\documentclass[lettersize,journal]{IEEEtran}
\usepackage{amsmath,amsfonts}
\usepackage[caption=true,font=footnotesize,labelfont=sf,textfont=sf]{subfig}
\usepackage{textcomp}
\usepackage{stfloats}
\usepackage{url}
\usepackage{verbatim}
\usepackage{graphicx}
\usepackage{cite}
\usepackage[T1]{fontenc}
\usepackage{booktabs}
\usepackage{makecell}
\usepackage{multirow}
\usepackage{multicol}
\usepackage{threeparttable}
\usepackage[table]{xcolor}
\usepackage{hyperref}

\definecolor{darkgreen}{RGB}{0,128,0}
\definecolor{mediumgreen}{RGB}{34,139,34}
\definecolor{lightgreen}{RGB}{100,200,100}
\definecolor{lightred}{RGB}{240,128,128}
\definecolor{mediumred}{RGB}{225,0,0}
\definecolor{darkred}{RGB}{178,34,34}
\definecolor{neutralgray}{RGB}{192,192,192}

\newcommand{\relcellpair}[2]{%
  #1 
  &
  \edef\Improvement{\fpeval{round(#1-#2,3)}}%
  \ifdim \Improvement pt > 0.20pt
    \cellcolor{darkgreen!60} +\Improvement
  \else\ifdim \Improvement pt > 0.09999pt
    \cellcolor{mediumgreen!50} +\Improvement
  \else\ifdim \Improvement pt > 0.0001pt
    \cellcolor{lightgreen!50} +\Improvement
  \else\ifdim \Improvement pt < -0.20pt
    \cellcolor{darkred!50} \Improvement
  \else\ifdim \Improvement pt < -0.10pt
    \cellcolor{mediumred!30} \Improvement
  \else\ifdim \Improvement pt < -0.0001pt
    \cellcolor{lightred!30} \Improvement
  \else
    \cellcolor{neutralgray!30} 0
  \fi\fi\fi\fi\fi\fi
}

\newcommand{\colorboxlegend}[2]{\textcolor{#1}{\rule{1ex}{1ex}}~#2}

\begin{document}

\title{Improving the Identification of Real-world Malware's DNS Covert Channels Using Locality Sensitive Hashing}

\author{
    Pascal Ruffing,
    Denis Petrov,
    Sebastian Zillien,
    and Steffen Wendzel
    
    \thanks{Pascal Ruffing is with the Center for Research and Technology (ZFT), University of Applied Sciences Worms, 67549 Worms, Germany (e-mail: ruffing@hs-worms.de).}
    \thanks{Denis Petrov is with the Center for Research and Technology (ZFT), University of Applied Sciences Worms, 67549 Worms, Germany, and also with the Institute of Information Resource Management (IRM), Ulm University, 89081 Ulm, Germany (e-mail: petrov@hs-worms.de).}
    \thanks{Sebastian Zillien is with the Institute of Information Resource Management (IRM), Ulm University, 89081 Ulm, Germany (e-mail: sebastian.zillien@uni-ulm.de).}
    \thanks{Steffen Wendzel is with the Institute of Information Resource Management (IRM), Ulm University, 89081 Ulm, Germany (e-mail: steffen.wendzel@uni-ulm.de).}
    \thanks{Corresponding author: Steffen Wendzel.}
}

\markboth{PRE-PRINT (Nov-25, 2025) / THIS PAPER IS CURRENTLY UNDER REVIEW}{}


\maketitle

\begin{abstract}
Nowadays, malware increasingly uses DNS-based covert channels in order to evade detection and maintain stealthy communication with its command-and-control servers. While prior work has focused on detecting such activity, identifying specific malware families and their behaviors from captured network traffic remains challenging due to the variability of DNS. In this paper, we present the first application of Locality Sensitive Hashing to the detection and identification of real-world malware utilizing DNS covert channels. Our approach encodes DNS subdomain sequences into statistical similarity features that effectively capture anomalies indicative of malicious activity. Combined with a Random Forest classifier, our method achieves higher accuracy and reduced false positive rates than prior approaches, while demonstrating improved robustness and generalization to previously unseen or modified malware samples. We further demonstrate that our approach enables reliable classification of malware behavior (e.g., uploading or downloading of files), based solely on DNS subdomains.
\end{abstract}


\section{Introduction}

\IEEEPARstart{T}{he covert} communication of a malware is considered one of its most fundamental features in orchestrating the exchange of secret messages between the malware clients and the corresponding command-and-control (C2) servers. To this end, malware employs different forms of network covert channels, i.e., policy-breaking and usually stealthy communication channels, nested inside other protocols \cite{csur}.

Recent years have shown an increase in network covert channel-capable malware \cite{Strachanski:StegomalwareSurvey,NeverMindMalware}.
Among the numerous network protocols utilized for tunneling, DNS was found to be the most used protocol during the years 2019-2024 \cite{Strachanski:StegomalwareSurvey}, which underpins the need to investigate the detectability and identification of malware using such DNS covert channels.

While much of the prior research has focused on the \emph{detection} of such malware ~\cite{DNSDetectionQi,DNSDetectionLi,DNSDetectionTu}, recent work has started to additionally \emph{identify} the particular \emph{malware family}~\cite{GraphTunnel} and the \emph{actions} that it performs (e.g., idling, uploading or downloading data, etc.), based solely on their DNS traffic \cite{ARES25:DNS}. Whereas most prior approaches concentrated on the subdomain portion of the DNS requests, other related techniques, such as \textit{Domain Generation Algorithms} (DGAs), create unusual structures in the second level domain. This is done by dynamically generating pseudo-random domain names which can evade existent blocklists and increase the overall resilience of the malware. Although this work focuses on covert tunneling, the methodology we propose could also be relevant for detecting such anomalies.

In this paper, we improve over previous work 
dealing with the detection and identification of DNS-based malware. To this end, our approach utilizes Locality Sensitive Hashing (LSH). We selected LSH for its ability to preserve structural similarity in inputs, enabling the efficient extraction of distinctive encoding patterns introduced by malware. Our primary contributions are:
\begin{enumerate}
    \item We present the first application of Locality Sensitive Hashing for the detection of DNS tunneling and the identification of the underlying malware, including its operational behavior.
    \item We improve the detectability, identification and behavior categorization compared to prior methods.
    \item While we employ the public malware dataset of~\cite{ARES25:DNS}, we enhance it through the integration of additional datasets such as~\cite{GraphTunnel} and~\cite{Ziza2023}, thereby covering a broader set of malware families and tunneling tools than previous papers.
    \item Our method demonstrates strong generalization, showing robustness against previously unseen malware variants and altered configurations that are not covered by previous research.
\end{enumerate}

The remainder of this paper is structured as follows: Sect.~\ref{sec_fun_rel_works} provides background information on DNS covert channels and Locality Sensitive Hashing, and highlights previous approaches and their limitations. Sect.~\ref{sec_methodology} details our methodology, including feature engineering and model training processes. Experimental evaluation and detailed analysis of detection and identification results, including a comparative study, are presented in Sect.~\ref{sec_evaluation}. Sect.~\ref{sec_discussion} discusses the broader implications of our findings, limitations of the approach, and directions for future research. Finally, Sect.~\ref{sec_conclusion} concludes.

\section{Fundamentals and Related Work}
\label{sec_fun_rel_works}

\subsection{DNS Tunneling}
DNS is a critical and ubiquitous component of the Internet communication, which is widely trusted by network infrastructures and security appliances. This implicit trust renders it a compelling target for misuse. One such attack vector is DNS tunneling, a type of a covert channel that leverages legitimate protocols to transmit data. It functions by having covert payloads embedded within the DNS query and response messages, enabling unauthorized bidirectional communication with remote C2 servers. These tunnels facilitate a range of malicious operations, including data exfiltration and payload delivery, while blending into regular DNS activity.

A common tunneling strategy manipulates DNS subdomains to carry arbitrary data, typically encoded in e.g., Base64 to comply with DNS syntax rules. For example, a DNS query of the form \textit{SGVsbG8gV29ybGQ.example.com} encodes the string \textit{Hello World} in the subdomain. The authoritative DNS server for \textit{example.com}, which is controlled by a malicious actor, acts as the C2 server. It extracts and decodes the content of the subdomain(s) and crafts a fitting response with further instructions. This work specifically focuses on this type of DNS tunneling.

\subsection{Locality Sensitive Hashing}
Locality Sensitive Hashing represents a family of algorithms designed to efficiently approximate the similarity between data points in high-dimensional space~\cite{jafari2021surveylocalitysensitivehashing}. Unlike cryptographic hash functions, which emphasize on producing highly diverse outputs even for similar inputs, LSH algorithms aim to ensure that similar inputs produce similar hash values. For example, two DNS subdomains that differ only by a single character map to nearly identical hash values, enabling efficient comparison based on similarity scores. 

For this work, Nilsimsa was selected due to its focus on structural patterns in textual data and its robustness to small modifications. Originally developed for detecting near-duplicate spam messages~\cite{damianiOpenDigestbasedTechnique2004Nilsimsa}, Nilsimsa computes a 256-bit digest that encodes the distribution of trigrams (three-character sequences) within an input string.

Nilsimsa processes the input by sliding a window of width five across the string and extracting all overlapping trigrams. Each trigram is mapped to one of 256 indices using a fixed lookup table combined with a permutation function that distributes trigram signatures uniformly across the bit space. For every occurrence of a trigram, the corresponding counter in a 256-bucket accumulator is incremented. After all trigrams have been processed, the accumulator values are thresholded at the median: buckets with counts above the median are assigned a bit value of 1, and all remaining buckets a value of 0. The resulting 256-bit vector constitutes the Nilsimsa digest and reflects which trigram groups occur more frequently than average.

Similarity between two Nilsimsa digests is computed by comparing their bit patterns and counting the number of matching bit positions. The resulting Nilsimsa \emph{compare value} ranges from $-128$ to $128$ and approximates the structural overlap between the underlying strings. High scores indicate substantial overlap in frequently occurring trigrams, whereas low or negative scores reflect limited or inconsistent structural similarity.

\subsection{Related Work}
\label{sec_related_works}
\subsubsection{Locality Sensitive Hashing for Anomaly Detection}
Several works have explored the use of LSH in security applications, where it has been applied to network traffic and malware samples for tasks such as anomaly detection, classification and clustering. 

One example is \textit{GIPS}, proposed by Seo and Yoon~\cite{Seo-LSH}, a streaming intrusion detection framework for zero-day attacks. By employing MinHash-based sketches to group similar data in real time, \textit{GIPS} generates signature-groups rather than individual signatures, thereby improving detection accuracy and significantly reducing false positives compared to prior automatic signature-generation methods.

Charyyev and Gunes developed an anomaly detection system tailored to the typically stable and repetitive communication patterns of IoT network traffic~\cite{charyyevDetectingAnomalousIoT2020}. Their approach employs Nilsimsa to cluster similar flow patterns and detect deviations from these clusters as indicators of anomalous or potentially malicious behavior. In a subsequent work, they extended this approach to also identify the type of detected anomalies by introducing \textit{LSADI}, which integrates multiple LSH functions and a majority-voting mechanism~\cite{charyyev2024}. This system is able to identify specific attack types, such as volumetric flooding (e.g., \textit{Smurf}, \textit{Fraggle}), and IoT malware including \textit{Mirai}, \textit{Hajime} and \textit{IRCBot}.

Peiser \textit{et al.} introduced a technique for detecting JavaScript-based malware that leverages multiple LSH methods to encode samples into LSH signatures~\cite{peiser_javascript_2020}. These signatures are subsequently processed by a feed-forward neural network, allowing them to classify malicious samples even in the presence of obfuscation techniques or malware variants.

To address the challenge of identifying polymorphic and evolving malware, Azab \textit{et al.} employed LSH to facilitate scalable clustering of malware samples~\cite{Azab}. Their method applies TLSH to efficiently group similar malware samples into clusters, enabling the discovery of previously unknown variants within large datasets. By focusing on approximate similarity rather than exact matching, their technique improves robustness against minor modifications commonly used to evade signature-based detection.

\subsubsection{DNS Tunneling Detection}
Wang \textit{et al.} presented a comprehensive survey on DNS tunnel detection methods, offering a taxonomy that distinguishes between rule- and model-based approaches \cite{wangComprehensiveSurveyDNS2021}. Rule-based methods are divided into signature- and threshold-based techniques, where the former relies on identifying predefined patterns or signatures of known tunneling tools, and the latter employs statistical thresholds on traffic or payload features to detect deviations from normal behavior. While signature-based methods can achieve high accuracy on known threats, they suffer from poor generalization and are ineffective against previously unseen tools. In contrast, model-based methods address these limitations by learning generalized patterns from data using traditional machine learning or deep learning approaches, allowing them to detect unknown and evolving tunneling methods.

One model-based approach was proposed by Buczak \textit{et al.}, in which Random Forest classifiers were applied to detect DNS tunnels in PCAP data~\cite{buczak}. They extracted 52 traffic features, including domain name length, query types, and packet sizes, and reported detection accuracies ranging from 90\% to 99\%, depending on the dataset used. Similarly, Schüppen \textit{et al.} \cite{FANCI} found Random Forest classifiers to outperform Support Vector Machines in detecting malware using domain generation algorithms. Their model was trained on 21 features capturing structural, linguistic and statistical characteristics of domain names, achieving accuracies above 0.98 in most experiments. The comparison of the similarity between benign and malicious DNS requests has been explored by Machmeier and Heuveline \cite{DNS-Time-Warping}. The approach taken by them makes use of a k-Nearest-Neighbour predictor and reaches a detection score of 99.9\%.

Žiža \textit{et al.} developed a detection approach that combines label-based and packet-based features~\cite{Ziza2023}. The label-based features focus on the text characteristics of the individual DNS requests, while the packet-based features describe properties of network flows with the same source, such as inter-request timings. Both a Random Forest and an XGBoost classifier were evaluated on data recorded at a university campus. During the capture period, the authors executed the \textit{Iodine} and \textit{DNSExfiltrator} tools with variations in the used parameters, as well as a custom modification of \textit{DNSExfiltrator}. Using this setup, Žiža \textit{et al.} achieved a highly accurate detection with a score of over 99\%.

The detection algorithm presented by Gao \textit{et al.}~\cite{GraphTunnel} is composed of a more complex setup that turns the parsed DNS paths into a graph which is then utilized in the prediction. Their GraphTunnel framework is reported to achieve a perfect 100\% accuracy on the test set of multiple DNS tunneling tools, including \textit{Iodine} and \textit{DNSCat2}. Furthermore, they achieve a near perfect accuracy when evaluating the robustness of the network to traffic generated by unknown tools, with an overall score of 99.37\% on a completely unknown dataset. In addition to the detection, Gao \textit{et al.} cover the identification of the samples within their datasets. This step takes place after the detection has taken place and uses statistical metrics built upon their features. Varying between each tool, their identification results show an accuracy of at least 98\%, with \textit{Cobaltstrike} and \textit{dnspot} reaching a 100\% score.

The work done by Petrov \textit{et al.}~\cite{ARES25:DNS} serves as the main comparison of our research, due to the overall similarity of the two approaches. Their Domainator framework has shown that using similarity metrics calculated purely upon the subdomains of the DNS queries allows a simple Random Forest classification to distinguish between malicious and benign traffic. In addition to that, the authors trained two further classifiers -- one that can identify the malware family which produced the traffic, and one that aims to determine the behavior of the malware. While the detection and identification scores of the classifiers reached above 0.96, the false positive rate was observed to be about 1.2\% for the benign traffic. In a real network setting, this would lead to non-malicious requests being incorrectly marked as malicious, thus decreasing the viability of the system while increasing the workload.
 
\section{Methodology}
\label{sec_methodology}

The core idea of this work is to use LSH to reveal structural patterns in DNS subdomain strings that distinguish legitimate from malware activity. This approach is based on the observation that most legitimate DNS queries for a given domain typically follow predictable, low-variance naming conventions, whereas malware systematically encodes data into subdomains, producing more variable patterns. These variations tend to reduce similarity within sequences of queries targeting the same domain. We capture this distinction by applying LSH to the subdomain portion of DNS queries and measuring pairwise similarity within sliding query windows. 

By computing similarity scores for all subdomain pairs in a window and summarizing their distribution using statistical features (e.g., mean, variance, quantiles), we derive a compact representation of the DNS traffic in that window. This feature representation serves as input to a machine learning model that can classify traffic as benign or malicious, and furthermore identify the specific DNS-based malware activities responsible for the queries.

\subsection{Training Dataset}
\label{ssec:training_dataset}
Our classification approach uses a constructed dataset consisting of both benign and malicious DNS traffic. It serves as the foundation for training and testing the machine learning-based pipeline described in Sect.~\ref{ssec_pipeline}.

The benign portion of the dataset is derived from the publicly available dataset collected by Žiža \textit{et al.}~\cite{Ziza2023}, which was originally curated for the analysis of DNS exfiltration attacks under adversarial conditions. It was captured in a real-world network environment, specifically from the request logs of the primary DNS server of a national ISP in Serbia. This subset reflects representative DNS request patterns across a wide range of typical use cases and serves as a reliable baseline for normal traffic behavior.

Apart from the legitimate requests, we incorporated authentic malware-generated DNS tunneling traffic, provided by Petrov \textit{et al.}~\cite{ARES25:DNS}. This dataset was constructed using the isolated testbed infrastructure described by Zillien \textit{et al.}~\cite{Zillien:WoDiCoFTestbed}, wherein multiple DNS malware families were executed under controlled conditions to record real-world tunneling behaviors. The malicious traffic was captured as a set of PCAP files, each representing DNS traffic from a specific malware, configuration, and behavioral scenario. These scenarios cover four distinct types of actions performed by the malware: an initial \textit{handshake} phase for registration and channel setup, \textit{idle} phases with periodic keep-alive messages, and the \textit{download} and \textit{upload} of different files from or to the C2 server. To facilitate efficient feature extraction and handling during development and training, we parsed these PCAPs into a single CSV file, which serves as the primary input to the machine learning pipeline.

The recorded malware families implement a variety of covert communication techniques, which are outlined in the following descriptions. \textit{Symbiote}, a Linux malware, infects running processes via \textit{LD\_PRELOAD} and hides its network activity by hijacking eBPF-based packet filtering~\cite{Symbiote}. It supports two modes: (1) \textit{exfiltration} of \textit{SSH} and \textit{SCP} credentials by encoding the data into A-record queries, and (2) \textit{downloading} further payloads from its C2 server to execute additional malicious actions. Unlike the other malware families in our dataset, \textit{Symbiote} does not exhibit a handshake phase or recurring idle traffic; its communication is limited to direct exfiltration and download activities.

\textit{Symbiote-DNSCat2} is a customized version of the open-source tunneling tool \textit{DNSCat2}, modified by the same attack group for use in their campaigns~\cite{Symbiote}. Compared to the original, it disables encryption, restricts communication to the \textit{TXT} record type, and hardcodes the target domain \textit{git.bancodobrasil.dev}. Despite these changes, the tool's core functionality remains the same. In its original form, \textit{DNSCat2} is not inherently malicious and provides bidirectional DNS-based communication for file transfer and command execution. The dataset also contains \textit{Iodine}, another open-source tool that tunnels IP traffic over DNS to bypass network restrictions and firewall policies. Although neither tool qualifies as real-world malware in a strict sense, their documented role in malware campaigns justifies their inclusion in the dataset as representative DNS-based malware communication~\cite{tsvetkovvladimirTTPsCyberPartisans2025,CuttingEdgePart,RussianMilitaryCyber2024}. 

\textit{Saitama} is a malware that targeted the Jordanian government in 2022~\cite{Saitama}. Unlike the previously mentioned malware families, it rotates among three exfiltration domains to enhance stealth and evade detection. Further improving on these qualities, the malware has an overall low throughput, with average pauses of a minute between each sent request, and additional longer breaks dependent on the behavior. 

The last malware family included in the dataset is \textit{RogueRobin}, appearing in two variants~\cite{RogueRobin,RogueRobin-Ironnet}. The first \textit{PowerShell} version appends non-encoded victim and job IDs to the beginning of the queried domain, but uses a base64 encoding to obfuscate the actual data that is transferred. In contrast to that, the subsequent \textit{.NET} version introduces a custom encoding pipeline that first uses a hexadecimal encoding and then translates all numbers into letters using a pre-defined scheme. Both malware variants use a round robin rotation of the used domain names and resource records, adding another level of obfuscation similar to \textit{Saitama}. 

An overview of the dataset scale, broken down by malware family and action type, is provided in Tab.~\ref{tab:malware_request_counts}. The malicious-to-legitimate ratio of approximately 60/40 retains the full behavioral diversity of the malicious samples while maintaining a sufficiently large and representative benign baseline. Adding more benign samples was not considered, as legitimate DNS traffic shows largely repetitive structures, and initial experiments revealed no measurable performance gains. This trade-off supports generalization while keeping training efficient and focused on behaviorally diverse samples.

\begin{table}[ht]
\centering
\caption{Distribution of DNS requests per malware family and action type in the training dataset.}
\label{tab:malware_request_counts}
\resizebox{\linewidth}{!}{%
\begin{tabular}{l|c|cccc}
\toprule
\textbf{Malware Family} & \textbf{Total} & \textbf{Handshake} & \textbf{Idle} & \textbf{Download} & \textbf{Upload} \\
\midrule
Symbiote          & 7,322  & --  & --     & 7,746  & 576 \\
Symbiote-DNSCat2  & 73,584 & 68  & 43,186 & 30,131 & 199 \\
DNSCat2           & 80,520 & 50  & 42,686 & 37,553 & 231 \\
Iodine            & 49,276 & 109 & 42,967 & 6,065  & 135 \\
Saitama           & 2,407  & 76  & 70     & 717    & 1,544 \\
RogueRobin-PS     & 30,040 & 190 & 11,074 & 17,687 & 1,089 \\
RogueRobin-Net    & 30,413 & 210 & 9,299  & 15,708 & 5,196 \\
\midrule
\textbf{Total Malicious} & \textbf{274,562} & \textbf{703} & \textbf{149,282} & \textbf{115,607} & \textbf{8,970} \\
\textbf{Total Legitimate}        & \textbf{178,822} & -- & --  & --    & -- \\
\bottomrule
\end{tabular}
}
\end{table}

\subsection{Processing Pipeline}
\label{ssec_pipeline}

\begin{figure}[ht]
    \centering
    \includegraphics[width=1.0\linewidth]{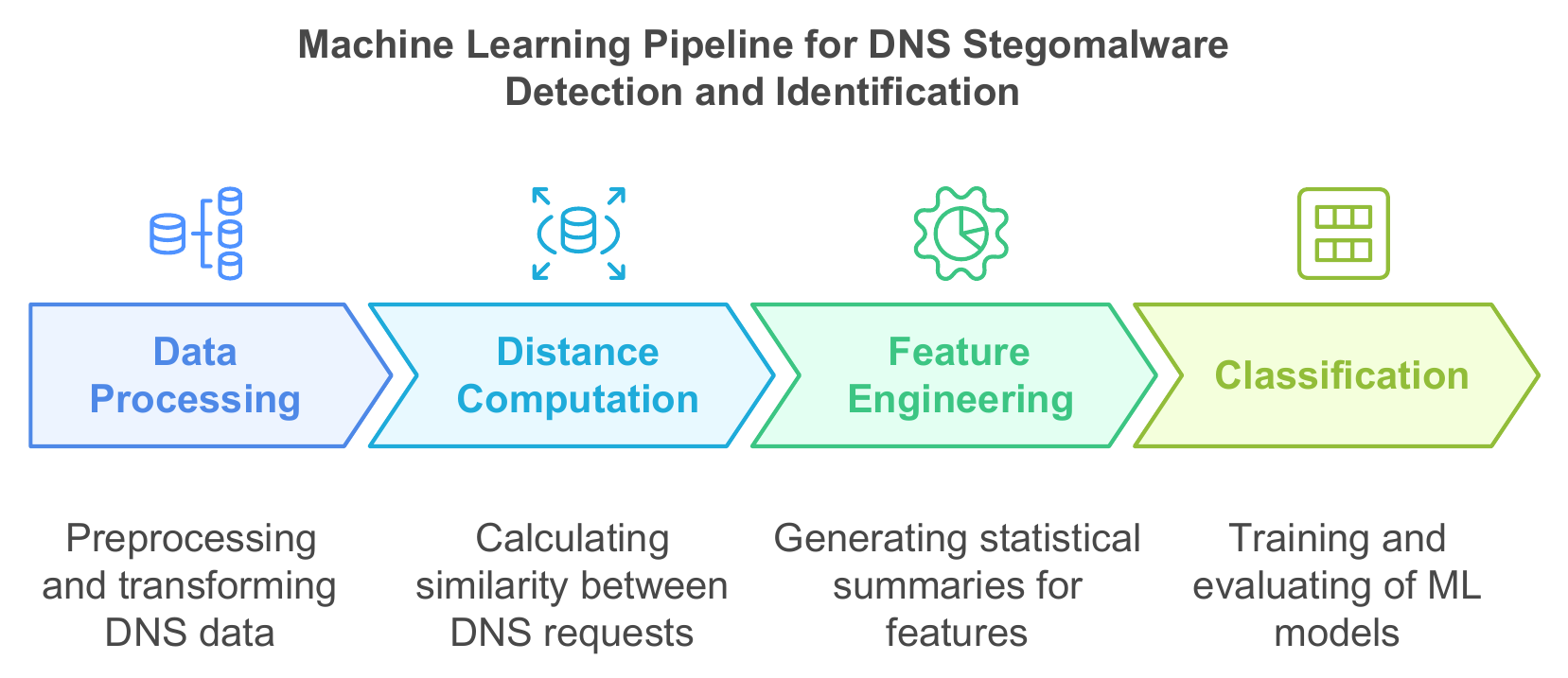}
    \caption{Overview of the machine learning pipeline for DNS tunneling detection and malware identification.}
    \label{fig:pipeline-overview}
\end{figure}

The proposed system follows a four-stage processing pipeline consisting of: (1) \textbf{Data Processing}, where subdomains are extracted, segmented, and transformed using Locality Sensitive Hashing; (2) \textbf{Distance Computation}, which performs pairwise comparisons within fixed-size windows; (3) \textbf{Feature Engineering}, where statistical summaries of similarity scores are computed for each segment; and (4) \textbf{Classification}, in which supervised machine learning models are trained to detect malicious traffic or identify specific malware families. This architecture is illustrated in Fig.~\ref{fig:pipeline-overview}, while the following subsections provide a detailed description of each stage.

\subsubsection{Data Processing} 
\label{subsec:data-processing}

In the initial processing stage, the subdomain portion of each DNS query is isolated by removing the registered domain and top-level domain. All delimiter characters are removed from the subdomain to eliminate artifacts that could bias subsequent similarity computations.

Each subdomain is then encoded through two complementary hashing strategies. First, the complete subdomain string is hashed in its entirety. Second, the string is partitioned into segments of approximately equal length, each of which is independently hashed. When the subdomain length is not evenly divisible by the number of segments, the remainder characters are distributed one by one starting from the first segment. This combined representation captures both global and localized characteristics of the query. 

After hashing, the DNS queries are grouped by their registered domain, consolidating them into a flow with the same target, while keeping the original order of the requests. This enables the model to analyze domain-specific behavior patterns and ensures that similarity computations reflect localized structural characteristics within coherent domain contexts.

\subsubsection{Distance Computation}
To capture temporal patterns in DNS traffic, the hashed queries are partitioned into non-overlapping windows of fixed size. This enables similarity analysis within a bounded context and supports both short- and long-term behavioral modeling, depending on the window size.

Within each window, all queries are compared in a pairwise fashion. The comparison itself is calculated between the corresponding subdomain segments, i.e., the first segment of each query is compared with first segment of every other query in the window. The same applies to the remaining segments. For a window of size \textit{n}, this results in $\binom{n}{2} = \frac{n(n - 1)}{2}$ comparisons per segment. This generates multiple sets of similarity scores per window, which are then passed to the next stage for statistical aggregation.

\subsubsection{Feature Engineering}
\label{ssec_feature_engineering}
The pairwise similarity scores within each window and segment are aggregated into a compact, uniform feature representation suitable for classification. For each segment we compute the following statistical descriptors: \textit{mean}, \textit{median}, \textit{first quartile (Q1)}, \textit{third quartile (Q3)}, \textit{variance}, \textit{minimum}, \textit{maximum}, and \textit{range}. These metrics are selected to capture both the central tendency and the spread of similarity values.

\begin{figure*}[t]
    \centering
    \subfloat[Mean vs. variance of pairwise LSH similarity scores.\label{fig:mean-vs-variance2}]{
        \includegraphics[width=0.48\textwidth]{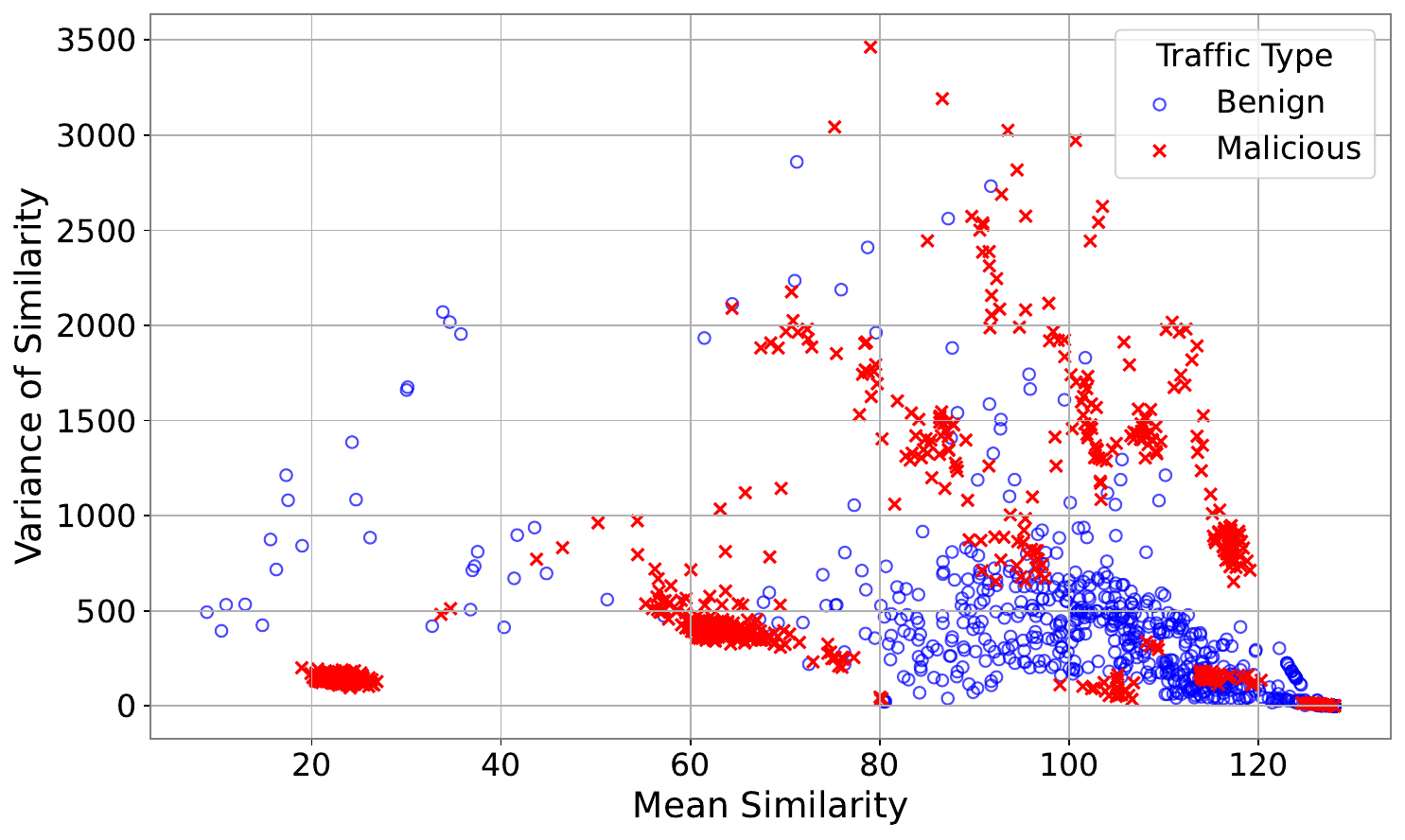}
    }\hfill
    \subfloat[Mean vs. range of pairwise LSH similarity scores.\label{fig:mean-vs-variance3}]{
        \includegraphics[width=0.48\textwidth]{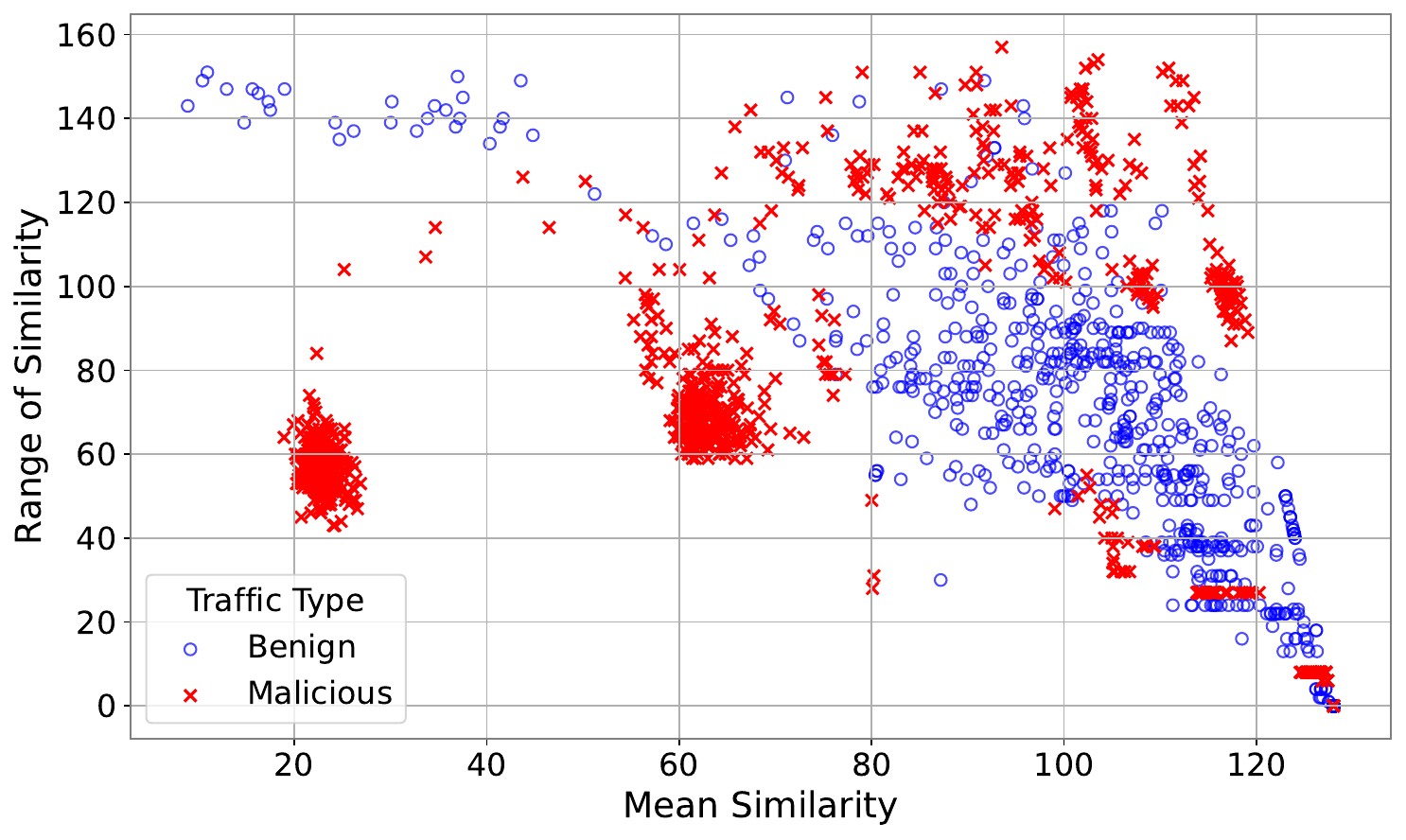}
    }
    \caption{Pairwise LSH similarity relationships for grouped windows of DNS requests from the \textit{Training Set}.}
    \label{fig:mean-vs-variance}
\end{figure*}

Fig.~\ref{fig:mean-vs-variance} illustrates how the feature set behaves under two selected feature combinations. Both exhibit clear structural differences between benign and malicious traffic, but with some regions of overlap that pose classification challenges. These overlaps are primarily caused by two specific patterns within the traffic. First, a subset of malicious windows exhibits unusually high mean similarity scores between 100 and 128 alongside low variance and range, primarily corresponding to idle communication phases. In these cases, such as observed for \textit{RogueRobin-PS}, identical queries are repeatedly issued to maintain the connection to its C2 server, which results in high similarity scores and causes these windows to closely resemble benign traffic. 

Conversely, certain benign windows show comparatively low mean similarity but high variance and range values. This behavior commonly occurs in DNS traffic from content-delivery networks, where dynamically generated subdomains produce frequent and non-repeating queries. As a result, these benign cases resemble the variability typically associated with malicious tunneling activity. When incorporating the full feature set, however, we are able to reduce these overlaps, resulting in clearer separation between classes.

\subsubsection{Classification}
In the final stage of the pipeline, supervised models are trained on the feature vectors extracted from each window, with labels indicating one of three classification tasks: \textbf{binary classification}, to distinguish malicious from benign traffic; \textbf{malware family classification}, to identify the specific DNS-utilizing malware; and \textbf{behavioral classification}, to determine the action performed by the malware.

Random Forest classifiers are employed for all classification tasks due to their robustness to feature variance and their effectiveness on small and imbalanced datasets.

\section{Evaluation}
\label{sec_evaluation}

We evaluate the proposed LSH-based pipeline using two complementary phases. The \textit{training phase} consists of the model training on the labeled dataset described in Sect.~\ref{ssec:training_dataset} using a 70/30 train-test split, and furthermore to validate key configuration choices of the pipeline (e.g., segmentation strategy, window size). For each classification task and window size, a dedicated model was trained and stored for subsequent use. In the \textit{evaluation phase}, these trained models are applied to previously unseen external datasets to assess their generalization performance.

\begin{table*}[ht]
\caption{Overview of the evaluation datasets, including the tools and malware families they contain, the number of extracted windows per window size, and the classification tasks used for evaluation.}
\label{tab:evaluation_datasets}
\centering
\begin{tabular}{l c l c c c c c c c}
\toprule
\textbf{Dataset} & \textbf{Year} & \textbf{Tools/Malware} & \multicolumn{6}{c}{\textbf{Number of Extracted Windows}} & \textbf{Covered Classifications} \\
\cmidrule(l){4-9}
& & & \textbf{5} & \textbf{10} & \textbf{20} & \textbf{30} & \textbf{40} & \textbf{50} & \\
\midrule
Variant Set & 2025 & \makecell[l]{Symbiote\\RogueRobin-PS/.NET\\Saitama\\DNSCat2\\Iodine} & 5,877 & 2,932 & 1,460 & 966 & 723 & 576 & \makecell[c]{Binary\\Family\\Behavioral} \\
\midrule
GraphTunnel-Known & 2024 & \makecell[l]{DNSCat2\\Iodine} & 37,568 & 18,784 & 9,392 & 6,261 & 4,696 & 3,756 & \makecell[c]{Binary\\Family} \\
\midrule
GraphTunnel-Unknown & 2024 & \makecell[l]{CobaltStrike\\dns2tcp\\OzymanDNS\\tcp-over-dns} & 94,888 & 47,422 & 23,709 & 15,805 & 11,853 & 9,482 & Binary \\
\midrule
Žiža-DNSExfil & 2023 & DNSExfiltrator & 9,451 & 4,725 & 2,362 & 1,575 & 1,181 & 945 & Binary \\
\midrule
Žiža-ModDNSExfil & 2023 & DNSExfiltrator & 8,460 & 4,230 & 2,115 & 1,410 & 1,057 & 846 & Binary \\
\midrule
Parssegny & 2025 & CobaltStrike & 7,931 & 3,965 & 1,982 & 1,321 & 990 & 792 & Binary
\\
\bottomrule
\end{tabular}
\end{table*}

\subsection{Evaluation Datasets}
\label{ssec_eval_datasets}
To enable the evaluation of our approach and its generalization capabilities across a broader range of configurations, tools and behaviors, we incorporated additional datasets that were not included in our training setup. 
While the dataset released by Petrov \textit{et al.}~\cite{ARES25:DNS} comprehensively captures real-world malware DNS traffic, no other such datasets were publicly available at the time of writing. To address this limitation, we introduced the datasets provided by Gao \textit{et al.}~\cite{GraphTunnel}, Žiža \textit{et al.}~\cite{Ziza2023} and Parssegny \textit{et al.}~\cite{Parssegny}. Although these datasets do not contain any real malware traffic, they include tunneling activity from widely utilized (open-source) DNS tunneling tools that are also employed in malicious campaigns~\cite{tsvetkovvladimirTTPsCyberPartisans2025,CuttingEdgePart,RussianMilitaryCyber2024}. This choice is motivated by the fact that DNS-based malware and generic DNS tunneling tools share fundamental operational characteristics: both encode arbitrary data into subdomain fields and transmit it via DNS queries to remote servers. Consequently, the structural patterns and behavioral signatures generated by these two classes of traffic are highly transferable, allowing the tunneling datasets to serve as a suitable proxy for the evaluation of our detection and classification pipeline.

The first dataset is the \textit{Variant Set} by Petrov \textit{et al.}, containing separate PCAP recordings of the same malware families as in the \textit{Training Set}. In order to add a level of variation, the samples were executed with altered configurations and communication characteristics. These modifications include uploading different files, altering domain names and resource records, adjusting the set of encoding schemes, and varying the size of the data transmitted per request. By simulating plausible adaptations of known threats, the dataset allows us to evaluate how well the classifier performs under realistic evasion scenarios that preserve the operational intent of the malware. To ensure a clear separation between training and evaluation, all altered samples were withheld from training, allowing us to assess model performance on structurally similar but unseen traffic.

The second and third datasets were sourced from the GraphTunnel project by Gao \textit{et al.}~\cite{GraphTunnel}. We separated this dataset into two subsets for our evaluation. The \textit{GraphTunnel-Known} subset includes DNS traffic generated by tunneling tools that were also present in our training data, namely \textit{DNSCat2} and \textit{Iodine}. Because no details about the performed actions are provided, this subset does not support behavior-level classification at the ground truth level and can only be used for detection and malware identification tasks. In contrast, the \textit{GraphTunnel-Unknown} subset contains traffic from four additional tools that were not included in our training data, and therefore represent previously unseen malware families from the perspective of the classifier. This subset is exclusively used for the detection task to evaluate the ability of the model to generalize to fully unknown tunneling tools and communication behaviors. 

Two further datasets are based on prior work by Žiža \textit{et al.}~\cite{Ziza2023}. \textit{Žiža-DNSExfil} includes various adversarial DNS exfiltration scenarios. For the evaluation, we utilize a subset containing \textit{DNSExfiltrator} traffic, excluding requests attributed to Iodine, as it consists only of minimal idle and handshake communication. \textit{Žiža-ModDNSExfil} is a modified version of \textit{DNSExfiltrator}, in which the authors introduced behavioral changes. Both datasets serve as an additional test case for the detection task, challenging the robustness of the classifier in the presence of unknown malware behaviors.

The last dataset upon which we evaluated our classifiers was published by Parssegny \textit{et al.}~\cite{Parssegny} and contained \textit{CobaltStrike} traffic, a remote access tool with DNS exfiltration capabilities, which has been widely used by malicious actors~\cite{CobaltStrike}. As this sample is not found within the training data, it serves as a way to further examine the detection generalization capabilities of our models.

An overview of all datasets, including their covered classification tasks and the number of malicious windows per window size, is provided in Tab.~\ref{tab:evaluation_datasets}. 

Several of the evaluation datasets contain only malicious traffic in their original form. To enable consistent and comparable analysis across all datasets, each malicious set was supplemented with an equal-sized subset of legitimate DNS queries, sampled from the same benign dataset described in Sect.~\ref{ssec:training_dataset}. To avoid data leakage, the benign samples used for evaluation were drawn from parts of the dataset not used during training. A separate random subset was selected for each evaluation dataset, ensuring that false positive rates reflect generalization to previously unseen legitimate traffic.

\subsection{Training Results and Segmentation Decision}
We first analyze how subdomain segmentation affects the LSH-based feature extraction pipeline. Specifically, we compared three strategies: no segmentation, splitting each subdomain into two segments, and splitting into three segments (cf. Sect.~\ref{subsec:data-processing}). All segmentation experiments were conducted within the \textit{training phase} using a 70/30 train-test split of the \textit{Training Dataset} described in Sect.~\ref{ssec:training_dataset}.

As shown in Tab.~\ref{tab:segmentation_accuracy}, all three configurations perform similarly on the \textit{test subset}, with two and three segments showing a slight edge over the ``no segmentation'' setup. However, these differences are small and inconsistent across window sizes and tasks, making it difficult to derive a clear decision based on training accuracy alone.

To resolve this, we further examined how well each of the setups performs when tasked with classifying unknown traffic behavior sourced from external datasets. The results diverge across tasks: for binary and family classification, two segments yield the highest accuracy and generalizes best. For example, on \textit{Žiža-ModDNSExfil}, classification accuracy dropped from 0.98 to 0.89 when using three segments instead of two.

Based on these findings, we adopted a task-specific segmentation strategy in all subsequent experiments: two segments were used for both binary and family classification, while three segments were used for behavioral classification.

\begin{table}[h]
\centering
\caption{Evaluation of segmentation strategies during the \textit{training phase}.}
\label{tab:segmentation_accuracy}

\subfloat[\textit{Binary Classification}]{
\begin{tabular}{c|c|c|c}
\toprule
\textbf{Window Size} & \textbf{No Segmentation} & \textbf{2 Segments} & \textbf{3 Segments} \\
\midrule
5  & 0.938 & 0.972 & 0.973 \\
10 & 0.958 & 0.972 & 0.973 \\
20 & 0.966 & 0.972 & 0.973 \\
30 & 0.971 & 0.974 & 0.975 \\
40 & 0.966 & 0.972 & 0.972 \\
50 & 0.971 & 0.974 & 0.974 \\
\bottomrule
\end{tabular}
}
\medskip

\subfloat[\textit{Malware Family Classification}]{
\begin{tabular}{c|c|c|c}
\toprule
\textbf{Window Size} & \textbf{No Segmentation} & \textbf{2 Segments} & \textbf{3 Segments} \\
\midrule
5  & 0.923 & 0.968 & 0.973 \\
10 & 0.954 & 0.971 & 0.973 \\
20 & 0.962 & 0.975 & 0.972 \\
30 & 0.968 & 0.969 & 0.974 \\
40 & 0.963 & 0.971 & 0.970 \\
50 & 0.967 & 0.973 & 0.974 \\
\bottomrule
\end{tabular}
}

\medskip

\subfloat[\textit{Behavioral Classification}]{
\begin{tabular}{c|c|c|c}
\toprule
\textbf{Window Size} & \textbf{No Segmentation} & \textbf{2 Segments} & \textbf{3 Segments} \\
\midrule
5  & 0.828 & 0.880 & 0.882 \\
10 & 0.864 & 0.888 & 0.895 \\
20 & 0.870 & 0.891 & 0.892 \\
30 & 0.883 & 0.888 & 0.895 \\
40 & 0.873 & 0.888 & 0.894 \\
50 & 0.887 & 0.902 & 0.906 \\
\bottomrule
\end{tabular}
}
\end{table}

\subsection{Binary Classification}
\label{ssec_detection}

\begin{figure}[ht]
    \centering
    \includegraphics[width=1.0\linewidth]{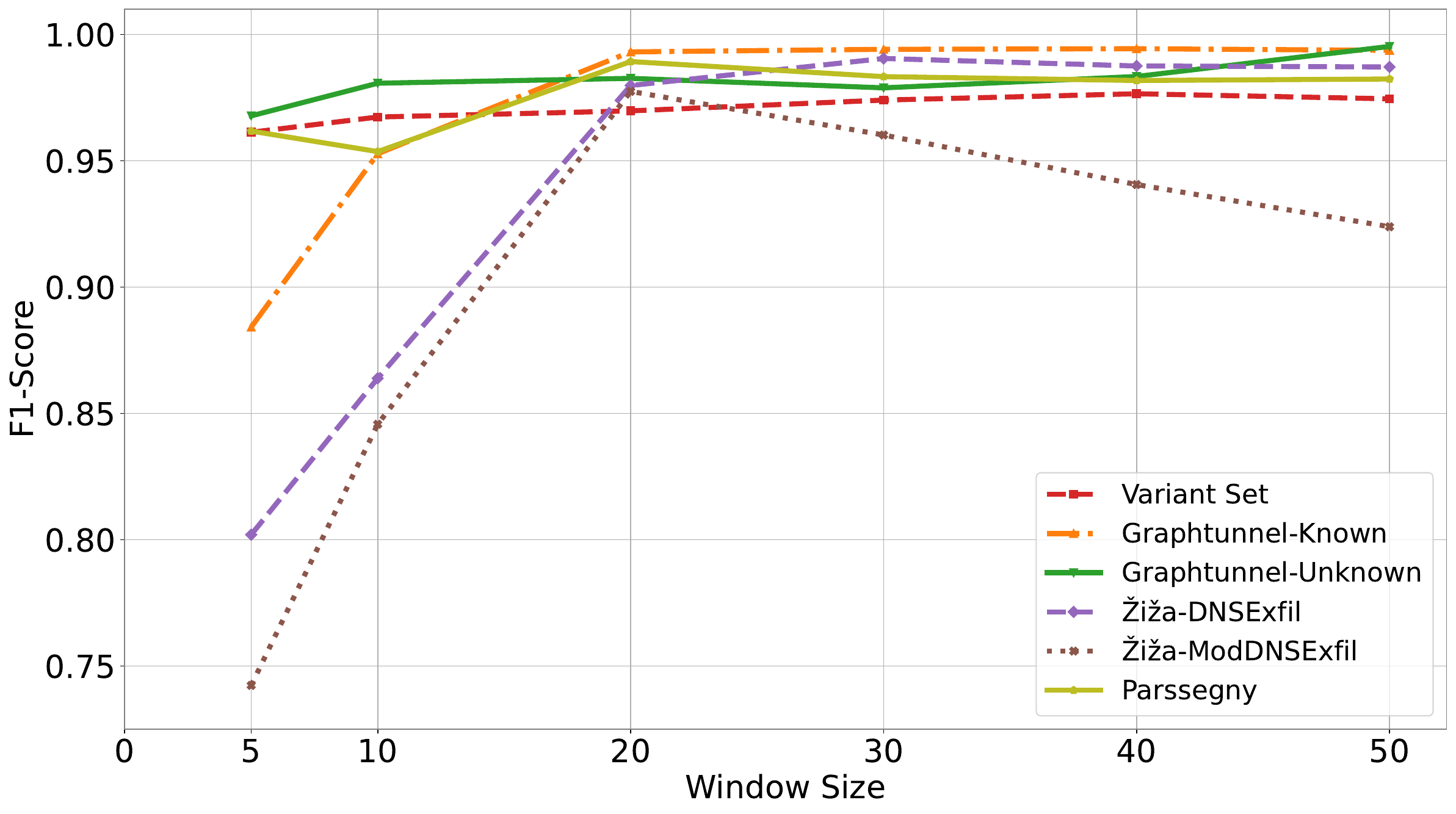}
    \caption{Binary classification across varying window sizes for all evaluated datasets.}
    \label{fig:detection_accuracy_plot}
\end{figure}

We first evaluate the ability of the pipeline to distinguish between malware and legitimate traffic. Fig.~\ref{fig:detection_accuracy_plot} presents the classification results for all evaluated datasets. Overall, the F1-score improves with increasing window size, with most datasets stabilizing around sizes 20-30. Beyond this, larger window sizes do not yield further improvements, likely due to over-smoothing effects from excessive aggregation. While the \textit{Variant} set maintain consistently high performance, the remaining datasets, particularly those involving previously unseen tunneling tools, benefit most from increased window sizes.

Beyond overall classification performance, we also considered the computational implications of varying the window size. Each DNS query requires LSH hashing of its subdomain segments, which grows linearly with window size $n$. More significantly, pairwise similarity comparisons grow quadratically, resulting in $\mathcal{O}(n^2)$ comparisons per window due to the cross-product between segment-wise hashes. This quadratic growth makes larger windows increasingly expensive in terms of runtime and memory consumption. 

While pairwise similarity computations scale quadratically with the window size $n$, this cost applies only to the initial batch. In a real-world deployment with a continuously incoming traffic stream, DNS requests arrive sequentially per domain. Using a rolling-window scheme, only the similarities between the newly arrived request and the existing $n-1$ entries need to be computed, while outdated values are discarded. This reduces the per-update complexity from $\mathcal{O}(n^2)$ to $\mathcal{O}(n)$ and enables efficient real-time processing. However, we did not employ this rolling-window mechanism in our evaluation setup, as the focus was on methodological comparability rather than deployment optimization. This aspect will be addressed in the future by extending the system toward real-time evaluation.

Although the F1-score remained consistently high across all tested window sizes, a window size of 20 was found to offer the best balance between detection quality and computational cost. In latency-sensitive applications such as real-time intrusion detection systems (IDS), such smaller windows are clearly preferable. Tab.~\ref{tab:window20_eval} summarizes the corresponding results for this setting.

\begin{table}[ht]
\centering
\caption{Binary classification performance for window size 20 across all evaluation datasets.}
\label{tab:window20_eval}
\resizebox{\linewidth}{!}{%
\begin{tabular}{l|c|c|c}
\toprule
\textbf{Dataset} & \textbf{F1-Score} & \textbf{False Positive Rate} & \textbf{Legitimate Windows} \\
\midrule
Variant Set         & 0.970 & 0.0031 & 2,241 \\
GraphTunnel-Known           & 0.993 & 0.0015 & 9,568 \\
GraphTunnel-Unknown         & 0.983 & 0.0035 & 24,259 \\
Žiža-DNSExfil                 & 0.980 & 0.0024 & 2,928 \\
Žiža-ModDNSExfil               & 0.977 & 0.0009 & 2,318 \\
Parssegny                     & 0.989 & 0.0005 & 1,835 \\
\midrule
\textbf{Average (Unweighted)} & \textbf{0.982} & \textbf{0.0020} & \textbf{---} \\
\textbf{Average (Weighted)}   & \textbf{0.984} & \textbf{0.0027} & \textbf{---} \\
\bottomrule
\end{tabular}
}
\end{table}

Tab.~\ref{tab:detection_accuracy} summarizes the classification results for selected evaluation datasets at a window size of 20, highlighting the distribution of correct and incorrect predictions. The remaining datasets, not included in Tab.~\ref{tab:detection_accuracy}, exhibit comparable patterns. 

Among the evaluated datasets, the \textit{Variant Set} shows the highest number of false negatives. A detailed breakdown of these misclassifications reveals that most originate from two malware families: \textit{RogueRobin-PS} (67 instances) and \textit{RogueRobin-Net} (28)--with additional cases observed for \textit{Iodine} and \textit{Symbiote} (2 each) and \textit{DNSCat2} (5). A closer examination of the misclassified \textit{RogueRobin-PS} instances shows that a majority originate from idle traffic. During this phase, the malware repeatedly sends identical DNS queries to request new tasks from its C2 server. Consequently, the statistical similarity features extracted from these windows exhibit virtually no variation, making them closely resembling benign traffic, which likewise often exhibits limited structural change across consecutive requests. This overlap in structural patterns causes the classifier to occasionally mislabel idle windows as legitimate.

\begin{table}[ht]
\centering
\caption{Confusion matrices for binary detection (window size~20) across selected evaluation datasets.}
\label{tab:detection_accuracy}

\subfloat[\textit{Variant Set}]{
    \begin{tabular}{lcc}
    \toprule
     & Pred.\ Legitimate & Pred.\ Malicious \\
    \midrule
    Actual Legitimate & \cellcolor{green!60}2{,}234 & \cellcolor{red!5}7 \\
    Actual Malicious  & \cellcolor{red!15}103 & \cellcolor{green!40}1{,}357 \\
    \bottomrule
    \end{tabular}
}
\vspace{5pt}

\subfloat[\textit{Žiža-ModDNSExfil}]{
    \begin{tabular}{lcc}
    \toprule
     & Pred.\ Legitimate & Pred.\ Malicious \\
    \midrule
    Actual Legitimate & \cellcolor{green!60}2{,}316 & \cellcolor{red!2}2 \\
    Actual Malicious  & \cellcolor{red!10}98 & \cellcolor{green!45}2{,}017 \\
    \bottomrule
    \end{tabular}
}
\vspace{5pt}

\subfloat[\textit{Parssegny}]{
    \begin{tabular}{lcc}
    \toprule
     & Pred.\ Legitimate & Pred.\ Malicious \\
    \midrule
    Actual Legitimate & \cellcolor{green!60}1{,}834 & \cellcolor{red!2}1 \\
    Actual Malicious  & \cellcolor{red!10}40 & \cellcolor{green!45}1{,}942 \\
    \bottomrule
    \end{tabular}
}
\end{table}

Overall, the results demonstrate that the model generalizes well across diverse evaluation settings, including behaviorally altered samples and previously unseen tunneling tools. The consistently high F1-scores across all datasets suggests that the statistical similarity features are not overfitted to specific malware types or configurations. Instead, they capture more fundamental structural deviations inherent to DNS tunneling. Even in adversarial scenarios such as \textit{Žiža-ModDNSExfil}, where evasion techniques explicitly target structural detection, the model remains robust.

\subsection{Malware Family Classification}
\label{ssec:indentification_eval}

Following the binary classification task, we evaluate the classifier’s ability to identify the specific malware responsible for each traffic window. This task represents a multi-class classification problem, where each window is assigned to one of the trained malware classes.

\begin{figure}[ht]
    \centering
    \includegraphics[width=1.0\linewidth]{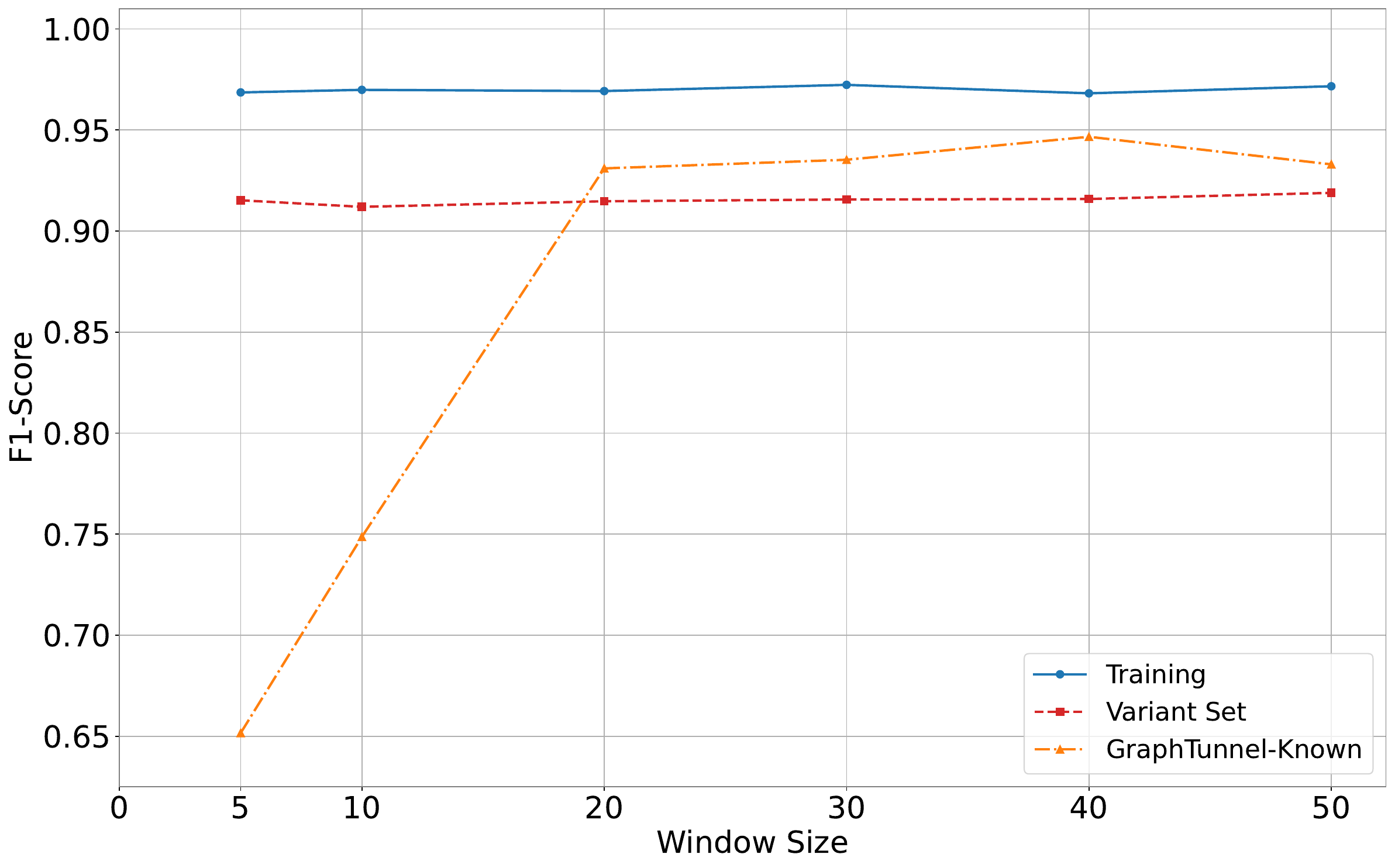}
    \caption{Family classification across varying window sizes.}
    \label{fig:identification_accuracy}
\end{figure}

Fig.~\ref{fig:identification_accuracy} shows the classification results for the \textit{Training}, the \textit{Variant Set}, and the \textit{GraphTunnel-Known} datasets. The classifier achieves consistently high F1-scores on the \textit{Training Set} and generally robust results for the \textit{Variant Set}, which includes behaviorally altered variants of the training malware. This classification task is inherently more difficult when performed on this dataset, as the modified malware exhibits similar but previously unseen structural patterns that were not included during training. This shifts the feature profiles of the trained classifier and leads to overlapping features between the samples, as behaviorally altered malware produces subdomain patterns that partially resemble multiple classes, reducing inter-class separability in the feature space. Overall, the results show that the model maintains a reliable performance, indicating a degree of robustness to behavioral variation. This robustness could likely be further improved in future works by incorporating a more diverse training dataset to better cover the range of possible protocol and configuration modifications.

\begin{figure}[ht]
\centering
\subfloat[\textit{Training}]{%
    \includegraphics[width=0.8\linewidth]{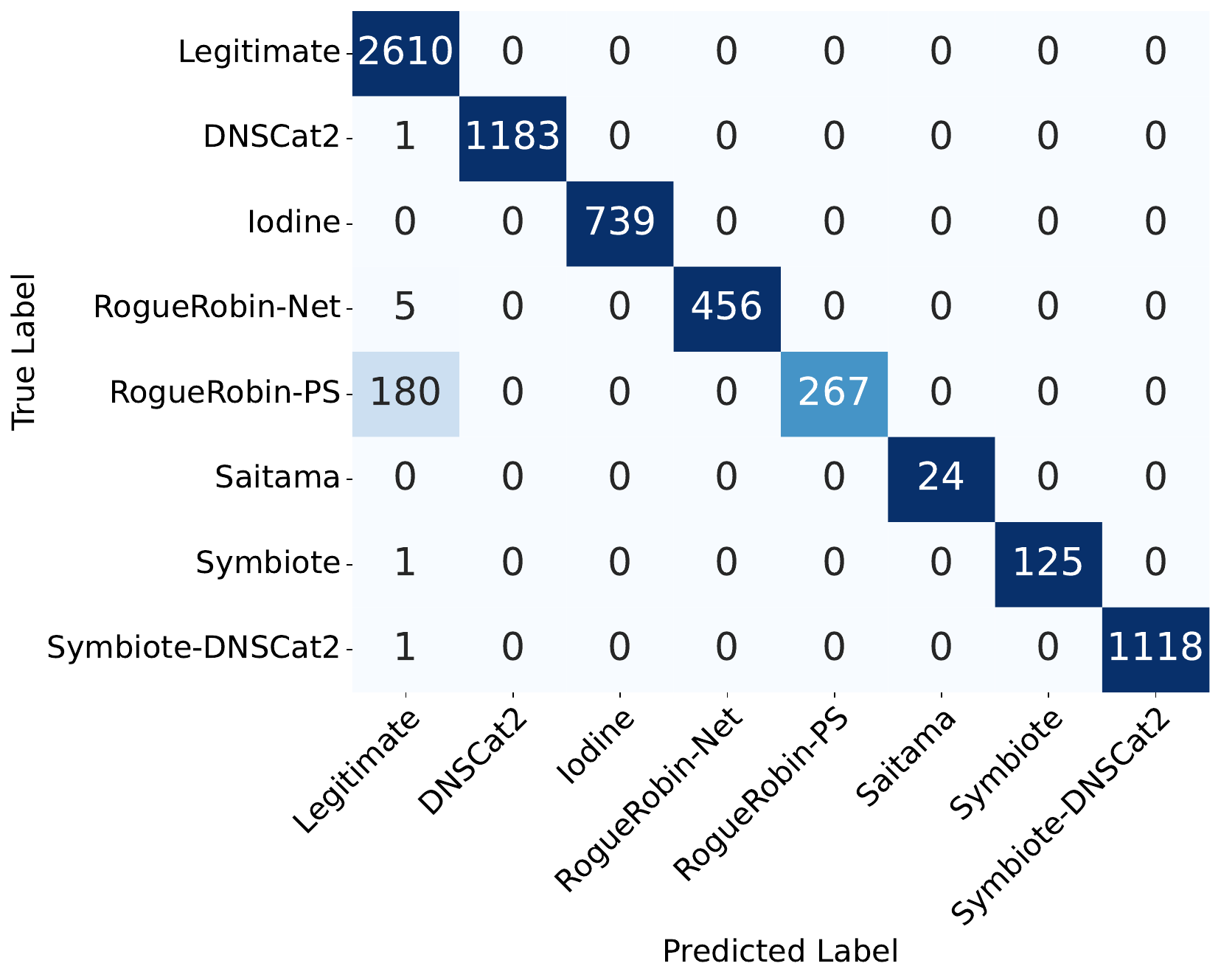}%
    \label{fig:ident_training_matrix}%
}
\hfill
\subfloat[\textit{Variant Set}]{%
    \includegraphics[width=0.8\linewidth]{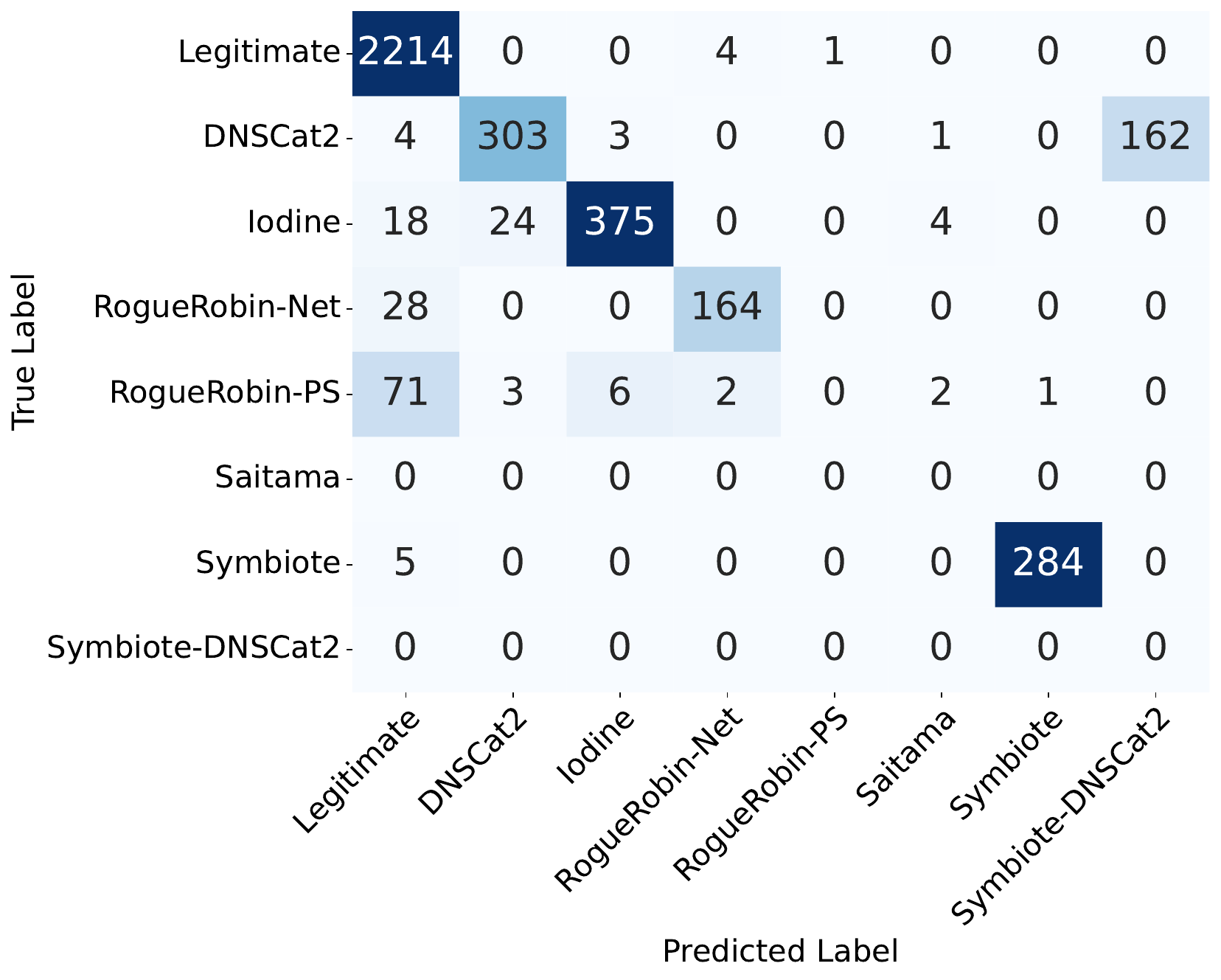}%
    \label{fig:ident_our_matrix}%
}


\subfloat[\textit{GraphTunnel-Known}]{%
    \includegraphics[width=0.8\linewidth]{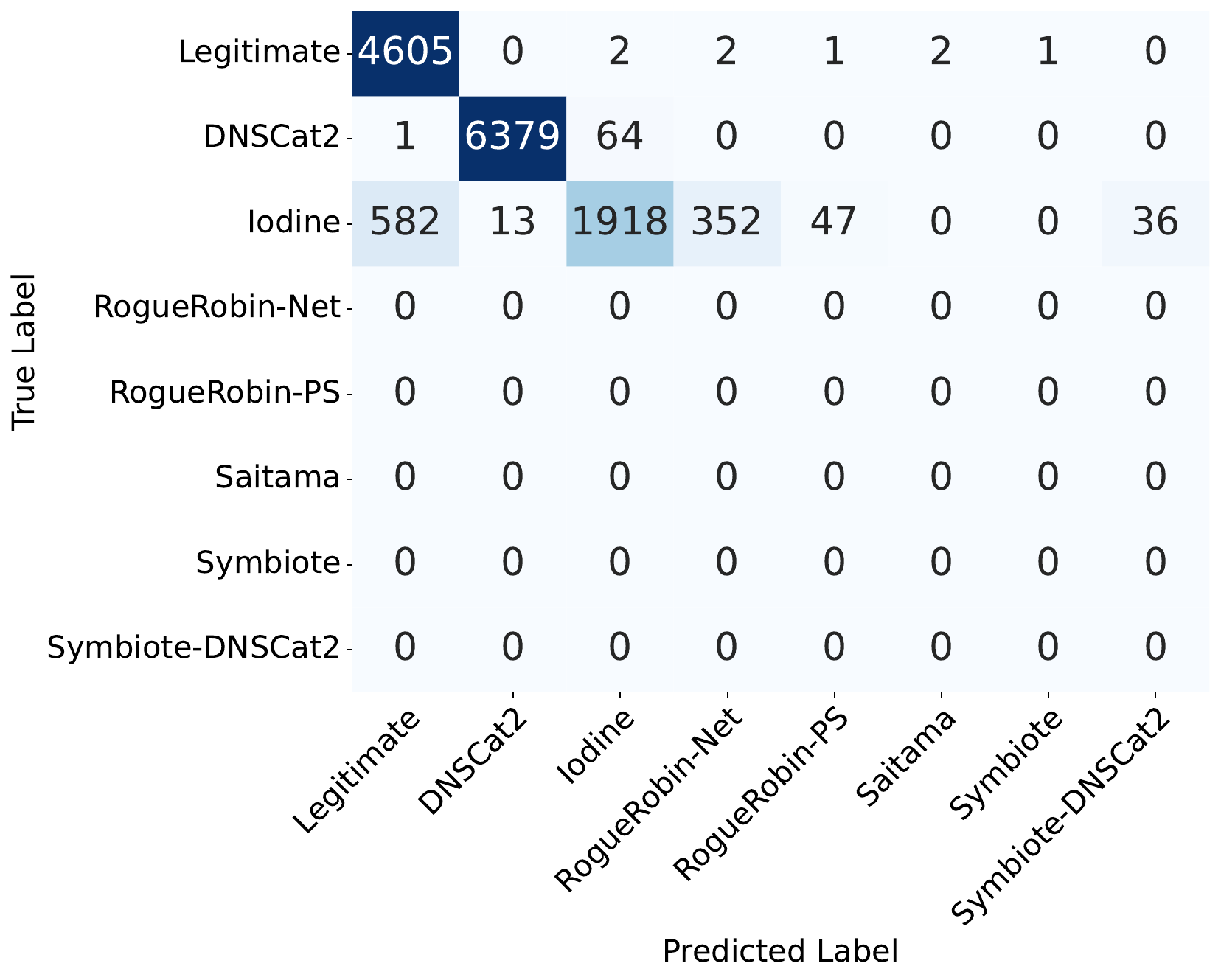}%
    \label{fig:ident_graph_matrix}%
}

\caption{Confusion matrices for the \textit{family} classification task at window size~20, across \textit{Training}, \textit{Variant Set}, and \textit{GraphTunnel-Known} datasets.}
\label{fig:confusion_matrices_identification}
\end{figure}

A detailed analysis of the performance is displayed in Fig.~\ref{fig:confusion_matrices_identification}, with all classifications produced using a window size of 20. The results from the \textit{training phase}, shown in Fig.~\ref{fig:ident_training_matrix}, serve as a reference for comparison. The classifier demonstrates high identification scores across all malware families and legitimate traffic. Particularly noteworthy is its ability to distinguish between \textit{DNSCat2} and \textit{Symbiote-DNSCat2}, despite both being based on the same underlying tool but configured with different runtime parameters. This suggests that the model has learned to recognize behavioral distinctions introduced by specific configurations, even when the core tunneling method itself is shared. However, this distinction is challenged on the \textit{Variant Set} displayed in Fig.~\ref{fig:ident_our_matrix}: here, modified \textit{DNSCat2} variants often resemble the configuration of \textit{Symbiote-DNSCat2}, leading to frequent misclassifications. These cases do not represent a failure of generalization but rather a labeling artifact: the traffic remains correctly associated with the broader \textit{DNSCat2} family, yet is assigned to the ``wrong'' class because the classifier treats some specific \textit{DNSCat2} configurations as \textit{Symbiote-DNSCat2}. As a result, the classifier appears less accurate than it actually is. To verify this, we repeated the experiments with the specific \textit{DNSCat2} configurations relabeled as \textit{Symbiote-DNSCat2}, which increased the overall F1-score from 0.91 to 0.94 and confirmed that the apparent errors were caused by the labeling scheme.

The classifier also performs poorly on \textit{RogueRobin-PS} variants, with none of the windows correctly classified and most mislabeled as legitimate. Unlike the \textit{Training Set}, where active communication patterns dominate, the \textit{Variant} set contains predominantly idle traffic that closely resemble benign DNS behavior. This behavior aligns with the binary classification results, where \textit{RogueRobin-PS} idle traffic likewise tends to be labeled as legitimate.

A possible improvement is to extend the classifier toward a hierarchical approach, first identifying the underlying malware family and subsequently distinguishing finer-grained runtime configurations or behavioral variants within that family. This would enable more precise attribution, even in the presence of variants that partially overlap with known classes. 

For the \textit{GraphTunnel-Known} dataset in Fig.~\ref{fig:ident_graph_matrix}, the classifier exhibits a failure mode similar to that observed for \textit{RogueRobin-PS} variants: while \textit{DNSCat2} and \textit{legitimate} traffic are separated with high reliability, \textit{Iodine} is frequently misclassified as legitimate or other malware families. This is largely attributable to the high proportion of idle traffic, accounting for approximately 50\% of the \textit{Iodine} samples, as well as the different runtime configurations used in this dataset. We examine the composition of this dataset in more detail in Sect.~\ref{ssec_behavior_graph}.

\subsection{Behavioral Classification}
To assess the ability of our system to classify malware not only by presence or type but also by its \textit{operational behavior}, we evaluate two distinct classification scenarios. These scenarios are designed to reflect different levels of contextual information and label granularity. In all scenarios, handshake traffic is excluded, as it typically contains only a small number of DNS queries, often representing a fixed and repetitive exchange pattern (e.g., registration or capability negotiation). Due to this low volume and structural uniformity, handshake traffic lacks sufficient variability and statistical richness to support meaningful classification.

\subsubsection{Compound Labels} 
The first scenario includes both legitimate and malicious DNS traffic. Malicious samples are labeled with a compound label encoding both the action and the malware family (e.g., \textit{Symbiote\_Upload}, \textit{Saitama\_Download}). This setting evaluates the model’s ability to distinguish malicious from benign traffic while also attributing the malicious behavior to the correct source. 

The best overall F1-score of 0.892 was achieved at a window size of 20, with the remaining window sizes yielding comparable scores. 12 out of 20 behavior-family combinations were classified with near-perfect F1-scores, including \textit{Iodine\_Upload}, \textit{Symbiote\_Download}, \textit{Saitama\_Upload}, and \textit{RogueRobin-Net\_Idle}, each achieving an F1-score of 1.0 or close to it. 

In contrast, the primary misclassifications were observed for \textit{RogueRobin\_PS\-\_Idle}, which was in almost all cases misclassified to legitimate traffic, a phenomenon we already observed in the binary and family classification results. The second major source of misclassification involved \textit{DNSCat2\_Idle} and \textit{DNSCat2\-\_Download}, which were frequently misclassified with each other due to their overlapping request patterns: during idle operation, the client repeatedly polls the server with empty data and changing sequence numbers, while in download mode the client issues similar polling requests that additionally include acknowledgments for previously received data chunks. Since both modes exhibit nearly identical characteristics, their statistical metrics overlap, which reduces separability between the two modes. 

\subsubsection{Action-Only Labels} 
The second scenario removes information tied to individual malware families from the labels. Each request is labeled solely by its operational intent (upload, download, idle, legitimate), irrespective of the originating source. By abstracting from malware identity, this setting focuses on recognizing general behavior classes, enabling detection of similar actions regardless of the specific malware families. 

\begin{figure}[ht]
    \centering
    \includegraphics[width=0.65\linewidth]{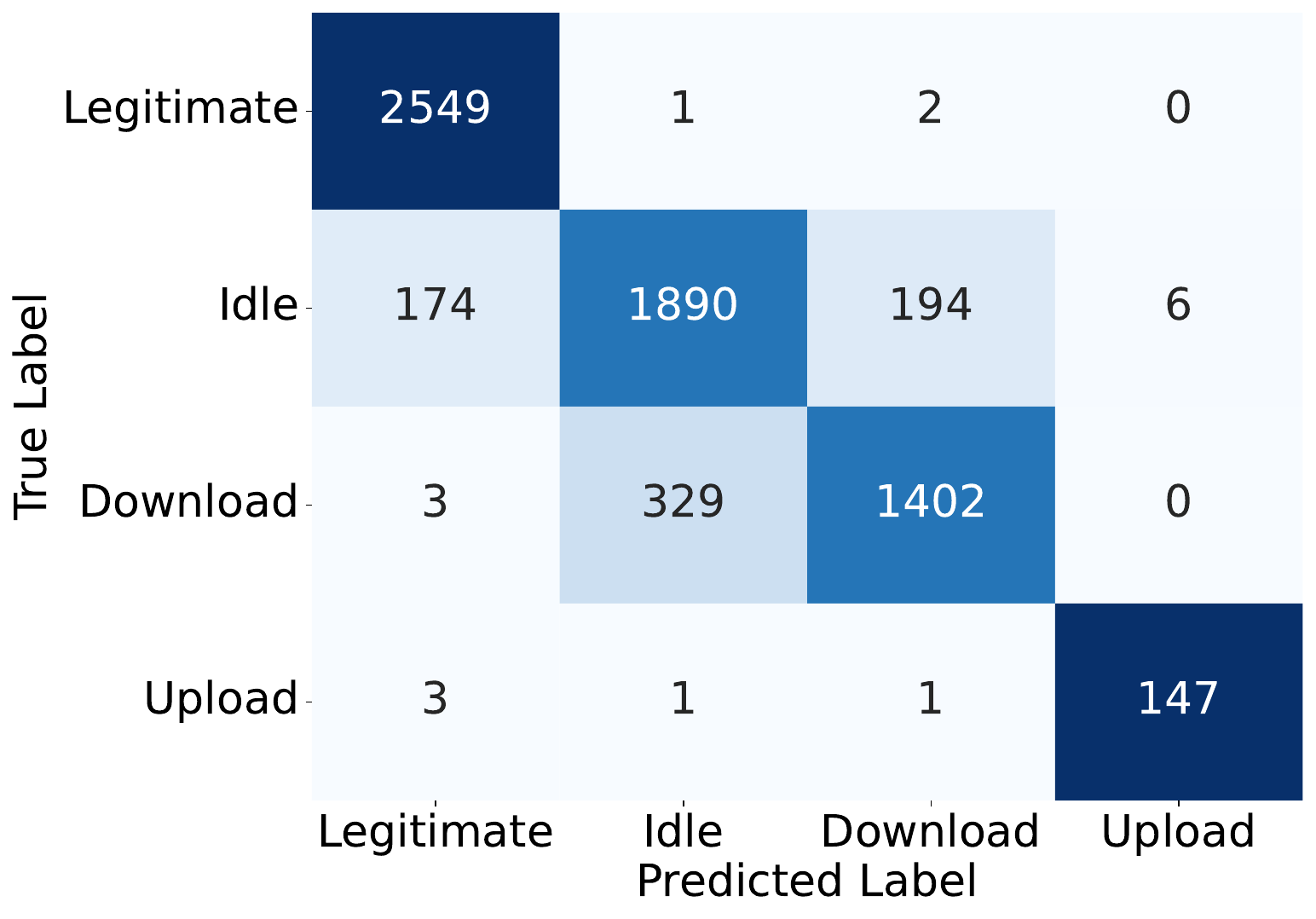}
    \caption{Confusion matrix for action-based traffic classification.}
    \label{fig:behaviour_matrix}
\end{figure}

The classifier achieves an overall F1-score of 0.893 at a window size of 20, closely matching the compound-label performance of the first scenario. As visualized in Fig.~\ref{fig:behaviour_matrix} , both upload and legitimate traffic are well-separated due to their distinct structural characteristics. Download traffic, while more variable, also remains sufficiently dissimilar from benign DNS behavior to allow robust discrimination in most cases. However, the dominant misclassification patterns occur between download and idle behaviors as observed in the first scenario.

\subsubsection{Examining \textit{GraphTunnel-Known} Data}
\label{ssec_behavior_graph}
As an additional evaluation step, we applied both labeling scenarios to the \textit{GraphTunnel-Known} dataset, despite the absence of action annotations by the authors.

For the action-only scenario, the predicted behavioral profile is dominated by upload activity, with \textit{DNSCat2} accounting for 6,435 windows and \textit{Iodine} for an additional 564. Download traffic is primarily associated with \textit{Iodine} (1,885 windows), while \textit{DNSCat2} contributes only nine such cases. A total of 494 \textit{Iodine} windows were classified as legitimate, even though the dataset itself contains no benign traffic. These cases therefore most likely correspond to idle activity, which aligns with the training results in Fig.~\ref{fig:behaviour_matrix}, where idle traffic is frequently misclassified as legitimate.

In contrast, the compound-label scenario produces a more fine-grained distribution, with predictions concentrated on specific tool–action pairs. The overall pattern shown in Tab.~\ref{tab:graph_action_pred} remains consistent with the action-only predictions, with \textit{DNSCat2} upload and \textit{Iodine} download dominating the recorded traffic. However, a subset of windows is classified as \textit{RogueRobin} uploads, despite this tool not being part of the dataset. A similar effect was already observed in the identification task (cf. Fig.~\ref{fig:ident_graph_matrix}), where a considerable number of \textit{Iodine} windows were mislabeled as \textit{RogueRobin-Net}.

\begin{table}[ht]
\centering
\caption{Predicted action distribution for \textit{GraphTunnel-Known}. Only predictions with at least 20 occurrences are shown.}
\label{tab:graph_action_pred}
\begin{tabular}{c|lr}
\toprule
\textbf{True Malware} & \textbf{Predicted Action} & \textbf{Count} \\
\midrule
\multirow{3}{*}{DNSCat2} & DNSCat2\_Upload & 5,584 \\
        & DNSCat2\_Download & 740 \\
        & DNSCat2\_Idle & 112 \\
\cmidrule(lr){1-3}
\multirow{3}{*}{Iodine}  & Iodine\_Download & 1,689 \\
 & Legitimate & 915 \\
 & RogueRobin-Net\_Upload & 299 \\
\bottomrule
\end{tabular}
\end{table}

\begin{table*}[t]
\caption{Performance comparison between Domainator and our approach for \textit{binary}, \textit{family}, and \textit{behavioral} classification tasks.}
\label{tab:domainator_comparison}
\centering
\begin{tabular}{llc|cc}
\toprule
\textbf{Task} & \textbf{Dataset} & \textbf{Metric} & \textbf{Domainator} & \textbf{Our Approach} \\
\midrule
\multirow{7}{*}{\rule{0pt}{12ex}\textbf{Binary Classification}}
& \multirow{2}{*}{Training}   & F1   & 0.966 & \textbf{0.972} \\
&                               & FPR  & 0.012 & \textbf{0.0008} \\
\cmidrule(lr){2-5}
& Variant Set           & F1   & --    & 0.970 \\
\cmidrule(lr){2-5}
& GraphTunnel-Known            & F1   & --    & 0.993 \\
\cmidrule(lr){2-5}
& GraphTunnel-Unknown          & F1   & --    & 0.983 \\
\cmidrule(lr){2-5}
& Žiža-DNSExfil                & F1   & 0.77  & \textbf{0.980} \\
\cmidrule(lr){2-5}
& Žiža-ModDNSExfil             & F1   & --    & 0.977 \\
\cmidrule(lr){2-5}
& Parssegny                     & F1   & --    & 0.989 \\
\midrule
\multirow{3}{*}{\rule{0pt}{4.5ex}\textbf{Family Classification}}
& Training                     & F1   & 0.968 & \textbf{0.969} \\
\cmidrule(lr){2-5}
& Variant Set          & F1   & --    & 0.915 \\
\cmidrule(lr){2-5}
& GraphTunnel-Known           & F1   & --    & 0.931 \\
\midrule
\multirow{2}{*}{\makecell[l]{\textbf{Behavioral Classification}\\Scenario 1}}
& Training                     & F1   & 0.857 & \textbf{0.892} \\
\cmidrule(lr){2-5}
& Variant Set          & F1   & --    & 0.824 \\
\midrule
\multirow{2}{*}{\makecell[l]{\textbf{Behavioral Classification}\\Scenario 2}}
& Training                     & F1   & 0.878 & \textbf{0.893} \\
\cmidrule(lr){2-5}
& Variant Set          & F1   & --    & 0.870 \\
\bottomrule
\end{tabular}
\end{table*}

\subsection{Comparative Evaluation with Domainator}

The primary goal of our evaluation is to demonstrate that the proposed LSH-based approach not only generalizes across diverse datasets but also improves upon the current state of the art. To this end, we conduct a direct comparison with the Domainator framework by Petrov \textit{et al.}~\cite{ARES25:DNS}, which represents the most comparable and reproducible baseline for DNS tunneling detection and identification. Prior works such as Žiža \textit{et al.}~\cite{Ziza2023} and Gao \textit{et al.}~\cite{GraphTunnel} provide strong evaluations but have not released their exact feature calculations or the models, preventing their inclusion in a fair benchmarking. Moreover, their scope is more restricted: Žiža \textit{et al.} focus exclusively on tunnel detection without identifying the underlying tool or behavior, while Gao \textit{et al.} does not address behavior analysis at all. 

As summarized in Tab.~\ref{tab:domainator_comparison}, our method achieves better \textit{detection} performance than Domainator, with consistently higher F1-scores and a significant reduction of false positive rate by 96\%. Domainator reports generalization results only for the \textit{Žiža-DNSExfil} dataset, where it reaches an F1-score of 0.77, while our approach achieves 0.98 on the same data. This suggests that the character-level feature design of Domainator fails to generalize to unseen tunneling behaviors, whereas our statistical representation delivers consistently high F1-scores and robustness across all evaluated datasets.

In the multiclass \textit{malware family classification} task shown in Tab.~\ref{tab:domainator_comparison}, our model shows slight improvements over Domainator on the \textit{training} dataset. For the \textit{Variant Set}, however, Domainator does not report an aggregated score. Instead, it performs a more isolated evaluation, reporting per-file results in which each source PCAP is treated separately and assigned its own F1-score. This differs from our own evaluation, where all recordings of a malware family were treated as a single stream, allowing windows to occasionally span transitions between different behaviors and thus reflecting a more realistic scenario. However, to enable a fair comparison, we replicated the approach by Petrov \textit{et al.} and computed corresponding per-file F1-scores, as shown in Tab~\ref{tab:per_file_results}. \textit{RogueRobin-PS} and \textit{Saitama} are not listed in this comparison, as their limited traffic volume does not permit a meaningful evaluation on individual PCAPs. All results were obtained using a two-step classification scenario: the family classification model is applied only to traffic that was previously classified as malicious, mirroring the evaluation design of Domainator.

Our model achieves perfect or near-perfect classification in 15 of the 21 evaluated PCAP files, compared to 5 of 21 for Domainator. Although these results demonstrate clear strengths, challenges remain in the malware family classification task as a whole. In particular, both approaches show similar misclassification patterns for certain \textit{DNSCat2} configurations. As outlined in Sect. \ref{ssec:indentification_eval}, these errors stem from the labeling scheme of the dataset, rather than limitations originating from of our classifier, as the comparable results of Domainator indicate that it is affected in the same way. 

\begin{table*}[h]
\caption{
Per-file F1-score comparison with Domainator for the family and behavioral classifications on the \textit{Variant Set}. Cell shading reflects the absolute F1-score differences (\(\Delta = \text{F1}_{\text{ours}} - \text{F1}_{\text{Domainator}}\)).
}
\label{tab:per_file_results}
\resizebox{\linewidth}{!}{%
\begin{threeparttable}
\centering
\begin{tabular}{c|c|c|c|c||c|c}
\toprule
\textbf{Tool} & \textbf{Scenario} & \textbf{Changes} & \multicolumn{2}{c||}{\textbf{Family Classification}} & \multicolumn{2}{c}{\textbf{Behavioral Classification}} \\
\cline{4-7}
 & & & \textbf{Our Approach} & \makecell{\textbf{Improvements} \\ \textbf{over Domainator}} & \textbf{Our Approach} & \makecell{\textbf{Improvements} \\ \textbf{over Domainator}}  \\
\midrule
\multirow{6}{*}{RR-Net} 
 & Upload & Different file & \relcellpair{1.0}{0.94} & \relcellpair{0.941}{0.90} \\
 & Upload & Data transfer length & \relcellpair{0.976}{0.74} & \relcellpair{0.320}{0.72} \\
 & Upload & Less domains & \relcellpair{1.0}{0.94} & \relcellpair{0.952}{0.83} \\
 & Upload & Changed encoding & \relcellpair{1.0}{0.90} & \relcellpair{1.0}{0.84} \\
 & Idle & Changed encoding & \relcellpair{1.0}{0.95} & \relcellpair{0.863}{0.93} \\
 & Download & Different file & \relcellpair{1.0}{0.91} & \relcellpair{0.868}{0.84} \\

\midrule
\multirow{7}{*}{Iodine} 
 & Upload & Data transfer length & \relcellpair{1.0}{0.90} & \relcellpair{1.0}{0.95} \\
 & Upload & Data transfer length & \relcellpair{0.240}{0.04} & \relcellpair{0.977}{0.0} \\
 & Idle & Data encoding & \relcellpair{0.995}{0.98} & \relcellpair{0.989}{0.98} \\
 & Idle & Resource Record TXT & \relcellpair{1.0}{0.94} & \relcellpair{0.995}{0.94} \\
 & Idle & Resource Record A & \relcellpair{1.0}{0.99} & \relcellpair{0.995}{0.99} \\
 & Download & Data transfer length & \relcellpair{1.0}{0.75} & \relcellpair{1.0}{0.75} \\
 & Download & Data transfer length & \relcellpair{1.0}{0.54} & \relcellpair{1.0}{0.50} \\

\midrule
\multirow{6}{*}{DNSCat2} 
 & Upload & Different file & \relcellpair{1.0}{0.94} & \relcellpair{0.970}{0.91} \\
 & Upload & Data encoding & \relcellpair{0.065}{0.03} & \relcellpair{0.235}{0.02} \\
 & Idle & Resource Record CNAME & \relcellpair{0.995}{1.0} & \relcellpair{0.754}{0.65} \\
 & Idle & Data encoding & \relcellpair{0.0}{0.0} & \relcellpair{0.993}{0.0} \\
 & Download & Different file & \relcellpair{1.0}{1.0} & \relcellpair{0.741}{0.4} \\
 & Download & Data encoding & \relcellpair{0.0}{0.0} & \relcellpair{1.0}{0.0} \\

\midrule
\multirow{2}{*}{Symbiote} 
& Upload & Different file & \relcellpair{0.923}{1.0} & \relcellpair{0.923}{1.0} \\
& Download & Different file & \relcellpair{1.0}{1.0} & \relcellpair{1.0}{1.0}  \\

\midrule
\multicolumn{3}{r|}{\textbf{Average improvement} (\(\overline{\Delta}\))} 
& \multicolumn{1}{c}{—} & \cellcolor{mediumgreen!50} \textbf{+0.08} 
& \multicolumn{1}{c}{—} & \cellcolor{mediumgreen!50} \textbf{+0.20} \\

\bottomrule
\end{tabular}
 \begin{tablenotes}
\small
\item \textbf{Color Legend:} 
\colorboxlegend{darkgreen!60}{strong improvement} (\( \Delta \geq 0.20 \)), 
\colorboxlegend{mediumgreen!50}{moderate improvement} (\( 0.10 \leq \Delta < 0.20 \)), 
\colorboxlegend{lightgreen!50}{weak improvement} (\( 0 < \Delta < 0.10 \)), 
\colorboxlegend{lightred!30}{weak decline} (\( -0.10 < \Delta < 0 \)), 
\colorboxlegend{mediumred!30}{moderate decline} (\( -0.20 < \Delta \leq -0.10 \)), 
\colorboxlegend{darkred!50}{strong decline} (\( \Delta \leq -0.20 \)), 
\colorboxlegend{neutralgray}{equal}.
\end{tablenotes}
\end{threeparttable}
}
\end{table*}

\begin{table}[ht]
\centering
\caption{Per-file family classification F1-scores for the \textit{GraphTunnel-Known} dataset compared to Domainator.}
\label{tab:graph_comparison}
\resizebox{\linewidth}{!}{%
\begin{tabular}{ll|c|c|c}
\toprule
\textbf{Tool} & \textbf{Configuration} & \textbf{Domainator} & \textbf{Our Approach} & \makecell{\textbf{Absolute Improvements} \\ \textbf{over Domainator}} \\
\midrule
\multirow{7}{*}{Iodine} 
  & A         & $\leq0.05$ & \relcellpair{0.969}{0.05} \\
  & CNAME     & $\leq0.05$ & \relcellpair{0.938}{0.05} \\
  & MX        & $\leq0.05$ & \relcellpair{0.900}{0.05} \\
  & SRV       & $\leq0.05$ & \relcellpair{0.969}{0.05} \\
  & NULL      & $\approx0.50$ & \relcellpair{0.617}{0.5} \\
  & TXT       & $\approx0.50$ & \relcellpair{0.529}{0.5} \\
  & Private   & $\approx0.50$ & \relcellpair{0.689}{0.5} \\
\midrule
DNSCat2 & {TXT, CNAME, MX} & 0.90 & \relcellpair{0.999}{0.90} \\
\midrule
\multicolumn{4}{r|}{\textbf{Average improvement} (\(\overline{\Delta}\))} & \cellcolor{darkgreen!60} \textbf{+0.50}\\
\bottomrule
\end{tabular}
}
\end{table}

Petrov \textit{et al.} also evaluated their classifier on the \textit{GraphTunnel-Known} dataset using a similar per-file breakdown and reported a mean F1-score of 0.90 across all \textit{DNSCat2} recordings. In contrast, our model achieves an F1-score of 1.0 on eight out of nine recordings, with the remaining one scoring 0.999. For \textit{Iodine}, the performance of Domainator varies considerably: according to their description, recordings using \textit{NULL}, \textit{TXT}, and \textit{private} resource records were identified approximately half of the time, while those using \textit{A}, \textit{MX}, \textit{CNAME}, and \textit{SRV} yielded F1-scores below 0.05. As shown in Tab.~\ref{tab:graph_comparison}, our approach performs substantially better across all configurations, particularly on those where Domainator consistently failed.

The \textit{behavioral classification} was then performed in a similar manner to the family classification, with these results also being displayed in Tab.~\ref{tab:per_file_results}. The overall performance for this task is notably higher than that of Domainator, which is ascertained by an average improvement of 20\%. Out of 21 files, 18 are equivalent or better than the state-of-the-art. From the few cases with decreased performance, only one has a reduction lower than 0.1 - the \textit{RogueRobin-Net} Upload with significant changes in the transfer length. This scenario stands out because the underlying queries have a median subdomain length of only nine characters, resulting in extremely short segments with limited statistical diversity for each. Domainator, by contrast, operates on the full subdomain string without segmentation, thereby retaining the full nine characters as a single unit and achieving a more robust performance for this specific configuration. For \textit{Iodine} configurations with reduced transfer length, the opposite effect occurs. Here, the advantages of segmentation are apparent: with a medium subdomain length of 79 characters, our method can extract more robust similarity features than Domainator, highlighting that segmentation becomes a strength once sufficient query length is available, in contrast to the short-query case of \textit{RogueRobin-Net}.

These findings suggest that adaptive mechanisms, for example dynamically adjusting the segmentation strategy, could further balance the trade-off between short and long queries, and represent a possible avenue for future work.

\section{Discussion}
\label{sec_discussion}
This study presents an LSH-based methodology for the analysis of DNS-based malware, introducing a novel contribution that leverages structural similarity features derived from DNS subdomain sequences. In contrast to prior approaches that rely on static signatures or extensive feature engineering, our method encodes pairwise similarity scores into statistical descriptors that generalize across malware families, runtime configurations, and encoding variants. The evaluation demonstrates the effectiveness of this representation across multiple classification tasks, including binary, family, and behavioral classification.

Compared to the Domainator framework, which also operates on the structural features of DNS traffic, our approach achieves consistent improvements across multiple evaluation dimensions. On the canonical train-test split, it reduces the false positive rate by 96\% in comparison to Domainator, from 1.2\% to 0.08\%, alongside an improvement in the F1-score. This improvement is especially relevant for real-world deployments, where false positives contribute significantly to operational overhead and can erode confidence in automated detection systems.

The model further demonstrates strong generalization, both to altered variants of known malware and to previously unseen tunneling tools. Configuration changes like different resource record types, or encoding logic, are typically handled robustly, with misclassifications occurring mostly within behaviorally adjacent classes. This abstraction extends to external datasets too: the model successfully detects covert DNS activity from six previously unseen tunneling tools without retraining.

Despite these strengths, limitations remain in traffic scenarios with low structural variance. Idle traffic from malware such as \textit{RogueRobin-PS} and \textit{Saitama} is frequently misclassified as benign, due to repetitive subdomain patterns that resemble legitimate DNS usage. Similarly, distinguishing between idle and download behavior in malware like \textit{DNSCat2} proves challenging, as both exhibit minimal intra-window variation and near-identical traffic. These edge cases highlight the limits of purely structure-based classification and point to opportunities for integrating temporal or contextual features in future work.

\section{Conclusion}
\label{sec_conclusion}

In this paper, we demonstrated that Locality Sensitive Hashing of DNS subdomains provides an effective foundation for detecting and identifying covert malware communication. Empirical results across multiple datasets confirm that, compared to prior approaches, our method achieves consistently higher F1-scores and markedly lower false positive rates, while maintaining strong generalization to both, behaviorally modified and previously unseen malware. It can reliably distinguish both malware families and operational behaviors. However, the observed limitation on handling low-variance idle traffic remain challenging and highlight the need to augment similarity with additional temporal or semantic features. Future work should therefore explore hybrid representations and extend our approach to other protocols. Beyond improving detection efficacy, these directions also hold promise for advancing forensic attribution.


\bibliographystyle{IEEEtran}
\bibliography{bibtex/IEEEabrv, sample-base}

@article{csur,
author = {Steffen Wendzel and Sebastian Zander and Bernhard Fechner and Christian Herdin},
title = {Pattern-Based Survey and Categorization of Network Covert Channel Techniques},
year = {2015},
issue_date = {April 2015},
publisher = {ACM},
volume = {47},
number = {3},
issn = {0360-0300},
doi = {10.1145/2684195},
journal = {ACM Computing Surveys},
articleno = {50},
numpages = {26}
}

@inproceedings{Zillien:WoDiCoFTestbed,
author = {Sebastian Zillien and Denis Petrov and Pascal Ruffing and Friedrich Gross},
title = {A Development Framework for {TCP/IP} Network Steganography Malware Detection},
year = {2024},
isbn = {9798400706370},
publisher = {ACM},
doi = {10.1145/3658664.3659651},
booktitle = {Proc.  IHMMSec},
pages = {95–100},
numpages = {6}
}

@article{GraphTunnel,
  author={Guangyuan Gao and Weina Niu and Jiacheng Gong and Dujuan Gu and Song Li and Mingxue Zhang and Xiaosong Zhang},
  journal={IEEE Trans. Information Forensics and Security},
  title={{GraphTunnel}: Robust {DNS} Tunnel Detection Based on {DNS} Recursive Resolution Graph},
  year={2024},
  doi={10.1109/TIFS.2024.3443596}
}

@inproceedings{Strachanski:StegomalwareSurvey,
author = {Fabian Strachanski and Denis Petrov and Tobias Schmidbauer and Steffen Wendzel},
title = {A Comprehensive Pattern-based Overview of Stegomalware},
year = {2024},
isbn = {9798400717185},
publisher = {ACM},
doi = {10.1145/3664476.3670886},
booktitle = {Proc. ARES '24},
}

@ARTICLE{NeverMindMalware,
  author={Luca Caviglione and Wojciech Mazurczyk},
  journal={IEEE Security \& Privacy},
  title={Never Mind the Malware, Here’s the Stegomalware},
  year={2022},
  volume={20},
  number={5},
  pages={101-106},
  doi={10.1109/MSEC.2022.3178205}
}

@misc{CobaltStrike,
  author = {Martin Sohn Christensen and Josh Abraham},
  title = {{Cobalt Strike}},
  url = {https://attack.mitre.org/software/S0154/},
  year = {2024}
}

@misc{RogueRobin,
  author = {{Palo Alto Networks}},
  title = {{RogueRobin}},
  note = {https://unit42.paloaltonetworks.com/unit42-new-threat-actor-group-darkhydrus-targets-middle-east-government/},
  year = {2018}
}

@misc{RogueRobin-Ironnet,
  author = {{Ironnet}},
  title = {{The siren song of RogueRobin}},
  url = {https://www.ironnet.com/blog/dns-tunneling-series-part-3-the-siren-song-of-roguerobin},
  year = {2020}
}

@misc{Saitama,
  author = {{Malwarebytes}},
  title = {{Saitama}},
  url = {https://www.threatdown.com/blog/apt34-targets-jordan-government-using-new-saitama-backdoor/},
  year = {2022}
}

@misc{Symbiote,
  author = {{Joakim Kennedy and The BlackBerry Threat Research \& Intelligence Team}},
  title = {{Symbiote Deep-Dive: Analysis of a New, Nearly-Impossible-to-Detect Linux Threat}},
  url = {https://intezer.com/blog/research/new-linux-threat-symbiote/},
  year = {2022}
}

@misc{Iodine,
  author = {{Erik Ekman and Iodine Contrib.}},
  title = {{iodine}},
  url = {https://github.com/yarrick/iodine},
  year = {2024}
}

@misc{DNSCat2,
  author = {{Ron Bowes and Contrib.}},
  title = {{DNSCat2}},
  url = {https://github.com/iagox86/dnscat2},
  year = {2024}
}

@Article{Ziza2023,
    author={Kristijan {\v{Z}}i{\v{z}}a
    and Predrag Tadi{\'{c}}
    and Pavle Vuleti{\'{c}}},
    title={{DNS} exfiltration detection in the presence of adversarial attacks and modified exfiltrator behaviour},
    journal={International Journal of Information Security},
    year={2023},
    volume={22},
    number={6},
    pages={1865-1880},
    issn={1615-5270},
    doi={10.1007/s10207-023-00723-w},
}

@inproceedings{buczak,
author = {Anna L. Buczak and Paul A. Hanke and George J. Cancro and Michael K. Toma and Lanier A. Watkins and Jeffrey S. Chavis},
title = {Detection of Tunnels in {PCAP} Data by Random Forests},
year = {2016},
publisher = {ACM},
doi = {10.1145/2897795.2897804},
booktitle = {Proc. Annual Cyber and Inf. Sec. Res. Conf. (CISRC '16')}
}

@INPROCEEDINGS{DNS-Time-Warping,
  author={Stefan Machmeier and Vincent Heuveline},
  booktitle={Cyber Security in Netw. Conf.},
  title={Detecting {DNS} Tunnelling and Data Exfiltration Using Dynamic Time Warping},
  year={2024},
  pages={83-91},
  doi={10.1109/CSNet64211.2024.10851475}
}

@inproceedings {FANCI,
    author = {Samuel Sch{\"u}ppen and Dominik Teubert and Patrick Herrmann and Ulrike Meyer},
    title = {{FANCI:} Feature-based Automated {NXDomain} Classification and Intelligence},
    booktitle = {Proc. USENIX Security '18},
    year = {2018},
    isbn = {978-1-939133-04-5},
    pages = {1165--1181},
    publisher = {USENIX Assoc.},
}

@inproceedings{ARES25:DNS,
    author = {Denis Petrov and Pascal Ruffing and Sebastian Zillien and Steffen Wendzel},
    title = {Domainator: Detecting and Identifying {DNS}-Tunneling Malware Using Metadata Sequences},
    booktitle = {Availability, Reliability and Security},
    year = {2025},
    series = {LNCS},
    publisher = {Springer Nature Switzerland},
    doi = {10.1007/978-3-032-00624-0_6},
    isbn = {978-3-032-00624-0},
    pages = {118--140}
}

@inproceedings{damianiOpenDigestbasedTechnique2004Nilsimsa,
  title = {An {{Open Digest-based Technique}} for {{Spam Detection}}},
  booktitle = {{{ISCA PDCS}}},
  author = {Ernesto Damiani and Sabrina Vimercati and Stefano Paraboschi and Pierangela Samarati},
  year = {2004},
  volume = {2004},
  pages = {559--564}
}

@misc{jafari2021surveylocalitysensitivehashing,
      title={A Survey on Locality Sensitive Hashing Algorithms and their Applications}, 
      author={Omid Jafari and Preeti Maurya and Parth Nagarkar and Khandker Mushfiqul Islam and Chidambaram Crushev},
      year={2021},
      eprint={2102.08942},
      archivePrefix={arXiv},
      primaryClass={cs.DB},
      url={https://arxiv.org/abs/2102.08942}, 
}

@inproceedings{charyyevDetectingAnomalousIoT2020,
  title = {Detecting Anomalous {IoT} Traffic Flow with Locality Sensitive Hashes},
  booktitle = {{{GLOBECOM}} 2020 - 2020 {{IEEE Global Communications Conference}}},
  author = {Batyr Charyyev and Mehmet Hadi Gunes},
  year = {2020},
  pages = {1--6},
  issn = {2576-6813},
  doi = {10.1109/GLOBECOM42002.2020.9322559}
}

@inproceedings{peiser_javascript_2020,
	location = {Cham},
	title = {{JavaScript} Malware Detection Using Locality Sensitive Hashing},
	isbn = {978-3-030-58201-2},
	doi = {10.1007/978-3-030-58201-2_10},
	pages = {143--154},
	booktitle = {{ICT} Systems Security and Privacy Protection},
	publisher = {Springer International Publishing},
	author = {Stefan Carl Peiser and Ludwig Friborg and Riccardo Scandariato},
	editor = {Hölbl, Marko and Rannenberg, Kai and Welzer, Tatjana},
	year = {2020},
	langid = {english},
}

@INPROCEEDINGS{Azab,
  author={Ahmad Azab and Robert Layton and Mamoun Alazab and Jonathan Oliver},
  booktitle={2014 Fifth Cybercrime and Trustworthy Computing Conference}, 
  title={Mining Malware to Detect Variants}, 
  year={2014},
  volume={},
  number={},
  pages={44-53},
  doi={10.1109/CTC.2014.11}}

@article{wangComprehensiveSurveyDNS2021,
  title = {A Comprehensive Survey on {{DNS}} Tunnel Detection},
  author = {Yue Wang and Anmin Zhou and Shan Liao and Rongfeng Zheng and Rong Hu and Lei Zhang},
  year = {2021},
  journal = {Computer Networks},
  volume = {197},
  pages = {108322},
  issn = {1389-1286},
  doi = {10.1016/j.comnet.2021.108322}
}

@ARTICLE{charyyev2024,
  author = {Batyr Charyyev and Mehmet Hadi Gunes},
  journal={IEEE Access}, 
  title={Identifying Anomaly in {IoT} Traffic Flow With Locality Sensitive Hashes}, 
  year={2024},
  volume={12},
  number={},
  pages={89467-89478},
  doi={10.1109/ACCESS.2024.3420238}}

@misc{tsvetkovvladimirTTPsCyberPartisans2025,
  title = {{TTPs} of {{Cyber Partisans}} Activity Aimed at Espionage and Disruption},
  author = {{Kaspersky ICS CERT}},
  year = {2025},
  langid = {american}
}

@misc{CuttingEdgePart,
  title = {Cutting {{Edge}}, {{Part}} 2: {{Investigating Ivanti Connect Secure VPN Zero-Day Exploitation}}},
  author = {Matt Lin and Robert Wallace and John Wolfram and Dimiter Andonov and Tyler Mclellan},
  journal = {Google Cloud Blog},
  url = {https://cloud.google.com/blog/topics/threat-intelligence/investigating-ivanti-zero-day-exploitation},
  langid = {english}
}

@misc{RussianMilitaryCyber2024,
  title = {Russian {{Military Cyber Actors Target US}} and {{Global Critical Infrastructure}}},
  year = {2024},
  author = {CISA},
  url = {https://www.cisa.gov/news-events/cybersecurity-advisories/aa24-249a},
  langid = {english}
}

@InProceedings{Parssegny,
author="Cl{\'e}ment Parssegny
and Johan Mazel
and Olivier Levillain
and Pierre Chifflier",
editor="Dalla Preda, Mila
and Schrittwieser, Sebastian
and Naessens, Vincent
and De Sutter, Bjorn",
title="Striking Back at Cobalt: Using Network Traffic Metadata to Detect Cobalt Strike Masquerading Command and Control Channels",
booktitle="Availability, Reliability and Security",
year="2025",
publisher="Springer Nature Switzerland",
address="Cham",
pages="163--185",
isbn="978-3-032-00624-0",
doi={10.1007/978-3-032-00624-0_8}
}

@inproceedings{Seo-LSH,
author = {HyungBin Seo and MyungKeun Yoon},
title = {Generative intrusion detection and prevention on data stream},
year = {2023},
isbn = {978-1-939133-37-3},
publisher = {USENIX Association},
address = {USA},
booktitle = {Proceedings of the 32nd USENIX Conference on Security Symposium},
articleno = {242},
numpages = {17},
location = {Anaheim, CA, USA},
series = {SEC '23},
url = {https://www.usenix.org/conference/usenixsecurity23/presentation/seo},
}

@Article{DNSDetectionLi,
AUTHOR = {Xinyu Li and Xiaoying Wang and Guoqing Yang and Jinsha Zhang and Chunhui Li and Fangfang Cui and Ruize Gu},
TITLE = {An Improved Approach to DNS Covert Channel Detection Based on DBM-ENSec},
JOURNAL = {Future Internet},
VOLUME = {17},
YEAR = {2025},
NUMBER = {7},
ARTICLE-NUMBER = {319},
ISSN = {1999-5903},
DOI = {10.3390/fi17070319}
}

@article{DNSDetectionTu,
  title = {{{DNS}} Tunnelling Detection by Fusing Encoding Feature and Behavioral Feature},
  author = {Yu Tu and Shuang Liu and Qian Sun},
  year = {2023},
  journal = {Computers \& Security},
  volume = {132},
  pages = {103357},
  issn = {0167-4048},
  doi = {10.1016/j.cose.2023.103357}
}

@article{DNSDetectionQi,
  title = {A Bigram Based Real Time {{DNS}} Tunnel Detection Approach},
  author = {Cheng Qi and Xiaojun Chen and Cui Xu and Jinqiao Shi and Peipeng Liu},
  year = {2013},
  journal = {Procedia Computer Science},
  volume = {17},
  pages = {852--860},
  issn = {1877-0509},
  doi = {10.1016/j.procs.2013.05.109}
}

\end{document}